\documentclass[twocolumn]{revtex4}
\usepackage[dvips]{graphicx}
\usepackage{float, color,array, multirow}

\begin{document}

\title{Spin ordering induced by lattice distortions in classical Heisenberg antiferromagnets on the breathing pyrochlore lattice}

\author{Kazushi Aoyama and Hikaru Kawamura}

\date{\today}

\affiliation{Department of Earth and Space Science, Graduate School of Science, Osaka University, Osaka 560-0043, Japan
}

\begin{abstract}
We theoretically investigate effects of lattice distortions on the spin ordering of classical Heisenberg antiferromagnets on the breathing pyrochlore lattice. In the model, local lattice distortions originating from the site vibration are taken into account to yield effective spin interactions via the spin-lattice coupling (SLC). The breathing lattice-alternation is characterized by the ratio of the nearest-neighbor interaction for large tetrahedra to that for small ones, $J'/J$. It is found by Monte Carlo simulations that the system exhibits a first-order transition into four different types of collinear magnetic ordered states. In the uniform case ($J'/J=1$), the state realized at stronger SLC is cubic-symmetric characterized by the $(\frac{1}{2},\frac{1}{2},\frac{1}{2})$ magnetic Bragg peaks, while the one at weaker SLC is tetragonal-symmetric characterized by the $(1,1,0)$ ones. With increasing the breathing distortion (decreasing $J'/J$), the ordering pattern of the $(\frac{1}{2},\frac{1}{2},\frac{1}{2})$ state becomes non-cubic with its magnetic Bragg reflections almost unchanged, while the $(1,1,0)$ state is robust. The non-cubic state peculiar to the breathing pyrochlores is further categorized into two types. We demonstrate that these two non-cubic orderings result in the massive degeneracy of the ground state and that the associated residual entropy per spin takes characteristic values of $\simeq \frac{\ln(4)}{16}k_{\rm B}$ and $\simeq \frac{\ln(6)}{16}k_{\rm B}$. Experimental implications of the results are discussed.
\end{abstract}

\maketitle
\section{introduction}
In frustrated magnets, it often happens that spins are coupled to the lattice degrees of freedom and the underlying lattice is distorted spontaneously to resolve the magnetic frustration, leading to a magnetic long-range order. The spinel chromium oxides ACr$_2$O$_4$ provide a typical example of such a spin-lattice-coupled ordering \cite{ZnCrO_Lee_00, CdCrO_Chung_05, HgCrO_Ueda_06, MgCrO_Ortega_08}, where the magnetic ion Cr$^{3+}$ forms the pyrochlore lattice, a three-dimensional network of corner-sharing tetrahedra. Of recent particular interest is another type of chromium oxides AA'Cr$_4$O$_8$ where Cr$^{3+}$ forms the so-called breathing pyrochlore lattice consisting of an alternating array of small and large tetrahedra. Roles of the breathing alternation on the spin-lattice-coupled ordering have been extensively studied in experiments \cite{BrPyro_Okamoto_13, BrPyro_Tanaka_14, BrPyro_Nilsen_15, BrPyro_Saha_16, BrPyro_Lee_16, BrPyro_Saha_17}. In view of such a situation, we theoretically investigate in this paper, effects of both the spontaneous distortion and the breathing alternation of the lattice on the spin ordering in pyrochlore antiferromagnets.         

In the uniform pyrochlore antiferromagnets ACr$_2$O$_4$ (A=Zn, Cd, Hg, Mg), each Cr$^{3+}$ ion has Hund-coupled three $3d$ electrons in the three-fold $t_{2g}$ level, so that it constitutes a $S=3/2$ localized spin. Since a magnetic anisotropy is relatively weak and the orbital degrees of freedom are absent, the classical Heisenberg model should provide a reasonable modeling. It is theoretically established that the classical Heisenberg spins on the pyrochlore lattice with the antiferromagnetic nearest-neighbor (NN) interaction do not order at any finite temperature due to a massive ground-state degeneracy \cite{Reimers_MC_92, Moessner-Chalker_prl, Moessner-Chalker_prb}. In real materials, weak perturbative interactions such as further-neighbor interactions would lift the degeneracy, eventually leading to a magnetic ordering or, alternatively, spins freeze in a highly-degenerate configuration leading to a residual macroscopic ground-state entropy.    
In the ACr$_2$O$_4$ compounds, the degeneracy is lifted by lattice distortions. It has been reported that the ACr$_2$O$_4$ family undergoes a first-order transition into a magnetic long-range-ordered state accompanied by a simultaneous structural transition which lowers the original cubic crystal symmetry to the tetragonal or orthorhombic one \cite{ZnCrO_Lee_00, CdCrO_Chung_05, HgCrO_Ueda_06, MgCrO_Ortega_08}. In spite of the spin-lattice-coupling (SLC) commonly seen in these compounds, the spin-ordering patterns vary from material to material \cite{ZnCrO_Lee_08, CdCrO_Chung_05, HgCrO_Matsuda_07, MgCrO_Ortega_08}.

Similar magnetostructural transitions have been also observed in the breathing pyrochlore antiferromagnets Li(Ga, In)Cr$_4$O$_8$ where due to the bond alternation, the NN interactions on small and large tetrahedra, $J$ and $J'$, take different values and the ratio $J'/J$ is estimated to be $J'/J \sim 0.1$ and $0.6$ in the In and Ga compounds, respectively \cite{BrPyro_Okamoto_13}. In LiInCr$_4$O$_8$ with stronger bond alternation, the magnetic transition splits up with the structural one and becomes of second order \cite{BrPyro_Tanaka_14, BrPyro_Nilsen_15, BrPyro_Saha_16, BrPyro_Lee_16}, whereas in LiGaCr$_4$O$_8$ with weaker bond-alternation, such a splitting is nominal and the transition remains of first order \cite{BrPyro_Tanaka_14, BrPyro_Nilsen_15, BrPyro_Saha_16}. In the both cases, the structural transition is incomplete and the low-temperature ordered phase is a coexistence of the original cubic and the emergent tetragonal crystal symmetries, each forming a distinct magnetic domain \cite{BrPyro_Nilsen_15, BrPyro_Saha_16}. Although extensive experimental studies have been done on compounds hosting breathing pyrochlore lattices such as substituted chromium spinel oxides \cite{BrPyro_Saha_17, BrPyro_doped_Okamoto_15, BrPyro_doped_Wang_17, BrPyro_doped_Wawrzynczak_17} and chromium spinel sulfides \cite{BrPyro_Sulfides_Okamoto_18, BrPyro_Sulfides_Pokharel_18}, relevant theoretical studies are limited to the Heisenberg model with only the bilinear NN interaction in which the lattice degrees of freedom are not taken into account \cite{BrPyro_NNmodel_Benton_15, BrPyro_NNmodel_Tsunetsugu_17, Pyro_NNmodel_Tsunetsugu_00}. With this NN model for the classical spin system, the massive ground-state degeneracy for the uniform pyrochlore lattice with $J'/J=1$ cannot be lifted by the change in the ratio $J'/J$ alone \cite{BrPyro_NNmodel_Benton_15}.
 
In this paper, to shed light on roles of the breathing structure on magnetic long-range orderings, we theoretically investigate effects of {\it local} lattice distortions on the ordering of classical Heisenberg spins in the presence of the breathing lattice-alternation, bearing the chromium oxides in our mind. Our analysis is based on the ``site-phonon'' Heisenberg model, which is a possible minimal model to describe the spin-lattice-coupled orders in the uniform pyrochlore antiferromagnets, as we will explain below.

Most of the theoretical studies on the SLC in the uniform pyrochlore lattice can be categorized into two main streams: one is a phenomenological theory \cite{SLC_Tchernyshyov_prl_02, SLC_Tchernyshyov_prb_02} based on a group theoretical approach developed by Yamashita and Ueda \cite{SLC_Yamashita_00}, and the other is a microscopic theory based on the so-called ``bond-phonon'' \cite{Bond_Penc_04, Bond_Motome_06, Bond_Shannon_10} and ``site-phonon'' \cite{Site_Jia_05, Site_Bergman_06, Site_Wang_08, Site_AK_16} Heisenberg models. The latter has bearing on this work. In the bond-phonon model, which was introduced to qualitatively explain the in-field properties of ACr$_2$O$_4$ \cite{ZnCrO_Miyata_jpsj_11, ZnCrO_Miyata_prl_11, ZnCrO_Miyata_jpsj_12, CdCrO_Kojima_08, HgCrO_Ueda_06}, the lattice deformation is assumed to occur independently at each {\it bond}, namely, independent bond-length vibrations are assumed. In reality, however, a magnetic ion at each {\it site} vibrates implying a strong correlation among the surrounding bond lengths. 

In the counter model of ``site phonon'', the Einstein model was assumed for the lattice-vibration part. In contrast to the ``bond-phonon'' model in which a first-order transition occurs into a nematic state with spins being collinearly aligned but not magnetically ordered unless perturbative interactions are incorporated \cite{Bond_Penc_04, Bond_Motome_06, Bond_Shannon_10}, the ``site-phonon'' coupling yields effective further neighbor interactions, inducing a first-order transition into two different types of collinear magnetic long-range orders. The state realized at stronger SLC is cubic-symmetric characterized by the magnetic  $(\frac{1}{2},\frac{1}{2},\frac{1}{2})$ Bragg peaks, while that at weaker SLC is tetragonal-symmetric characterized by the $(1,1,0)$ ones, each accompanied by commensurate local lattice distortions \cite{Site_AK_16}. Experimental data show that complex Bragg-peak patterns of ACr$_2$O$_4$ (A=Zn, Cd, Hg) \cite{ZnCrO_Lee_08, HgCrO_Matsuda_07, CdCrO_Chung_05} basically involve $(1,1,0)$ reflections as observed in the weak SLC regime, suggesting that the site-phonon coupling is a key ingredient for the magnetic ordering in these compounds. In this paper, we will examine how the above ordering properties of the site-phonon Heisenberg model are affected by the nonuniformity of the breathing lattice-distortion.

\begin{figure}[t]
\includegraphics[scale=0.65]{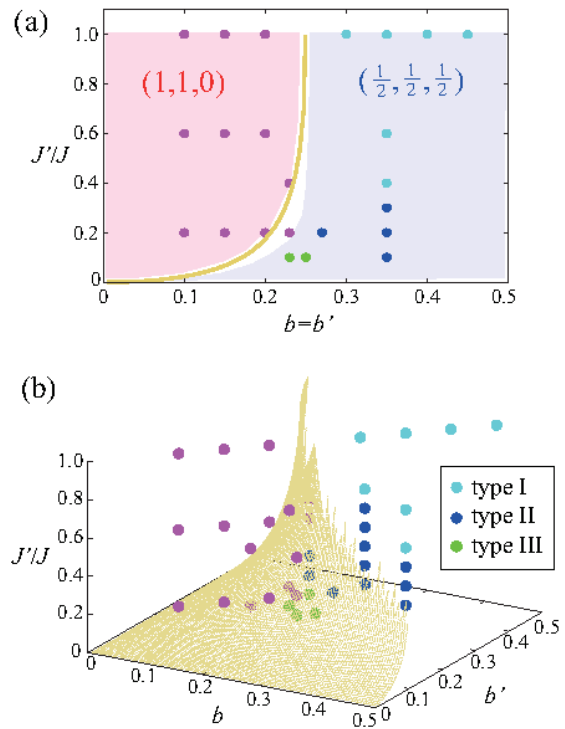}
\caption{The parameter dependences of the low-temperature ordered phases in the site-phonon Heisenberg model. There are four different types of collinear spin states. One is characterized by the $(1,1,0)$ magnetic Bragg peaks, and the others by the $(\frac{1}{2},\frac{1}{2},\frac{1}{2})$ ones. (a) Their stability regions in the $b$-$J'/J$ plane for $b'=b$, and (b) those in the $b$-$b'$-$J'/J$ space. Results of our MC simulations are represented by colored points. The tetragonal-symmetric $(1,1,0)$ state [Fig. \ref{fig:weak}(c)] is obtained at pink colored points, whereas the three $(\frac{1}{2},\frac{1}{2},\frac{1}{2})$ states, the cubic-symmetric type-I [Fig. \ref{fig:typeI}(d)], non-cubic type-II [Fig. \ref{fig:typeII}(d)], and non-cubic type-III [Fig. \ref{fig:typeIII}(d)] orders are obtained at cyan, blue, and green colored points, respectively. The orange curve and mesh denote the analytically obtained phase boundary between the $(1,1,0)$ and $(\frac{1}{2},\frac{1}{2},\frac{1}{2})$ states [Eq. (\ref{eq:boundary}) in (a) and Eq. (\ref{eq:boundary_general}) in (b)]. \label{fig:T0diagram}}
\end{figure}

 Our results are summarized in the phase diagram of Fig. \ref{fig:T0diagram}, where $J'/J$ and $b$ ($b'$) are dimensionless parameters measuring the breathing alternation and the SLC for small (large) tetrahedra, respectively. There are four different types of collinear spin-ordered states. One is characterized by the $(1,1,0)$ magnetic Bragg peaks, and the others by the $(\frac{1}{2},\frac{1}{2},\frac{1}{2})$ ones. The former (latter) is realized in the weak (strong) SLC regime with smaller (larger) values of $b$ and $b'$. Both the tetragonal-symmetric $(1,1,0)$ state and the cubic-symmetric $(\frac{1}{2},\frac{1}{2},\frac{1}{2})$ one appearing on the uniform pyrochlore lattice with $J'/J=1$ are robust against moderately breathing lattice-distortions, namely, moderate decrease in $J'/J$. With further decreasing $J'/J$, the originally cubic ordering pattern of the $(\frac{1}{2},\frac{1}{2},\frac{1}{2})$ state, which we call type I as will be detailed in Sec. V A, becomes non-cubic with its magnetic Bragg reflections almost unchanged, which we call type II and III as will be detailed in Sec. V B. We demonstrate that in the type-II and type-III orders, the non-cubic spin configuration possesses a massive ground-state degeneracy which manifests itself as a distinct macroscopic residual entropy associated with its ordering rule.

This paper is organized as follows: In Sec. II, we introduce the microscopic site-phonon Heisenberg model. The basic ordering properties of this model will be discussed based on analytical calculations in Sec. III. In Secs. IV and V, we will show results of our Monte Carlo (MC) simulations for the $(1,1,0)$ state in the weak SLC regime and the $(\frac{1}{2},\frac{1}{2},\frac{1}{2})$ states in the strong SLC regime, respectively. We end the paper with summary and discussions in Sec. VI.

\section{model}
We derive our spin Hamiltonian describing the SLC in the presence of the breathing lattice-distortion. Throughout this paper, the NN sites denote the neighboring sites connected by a single bond independent of its length, while further neighboring sites such as the second and third NN ones are defined in the same way as that for the uniform pyrochlore lattice, namely, by the distance between two sites. In the site-phonon model, the displacement vector ${\bf u}_i$ at each site $i$ from its regular position ${\bf r}^0_i$ on the lattice is assumed to be independent of the ones at the neighboring sites. Although in reality neighboring ${\bf u}_i$'s should be correlated to each other, we use here the site-phonon model because it is the simplest and minimal model describing phonon-mediated spin interactions. 
In addition, spin correlations mediated by the phonons can conversely influence the phonon properties via SLC \cite {ZnCrO_Sushkov_05, ZnCrO_Fennie_06, CdCrO_Rudolf_07, CdCrO_Aguilar_08, CdZn-CrO_Kant_09, CdCrO_Bhattacharjee_11}, but such a feedback effect is ignored in this model for simplicity. 

In the site-phonon picture, an appropriate minimal spin-lattice-coupled model might be
\begin{equation}\label{eq:original_H}
{\cal H} = \sum_{\langle i,j \rangle } J_{\rm ex}\big(|{\bf r}^0_{ij} + {\bf u}_i-{\bf u}_j|\big){\bf S}_i \cdot {\bf S}_j + \frac{c}{2}\sum_i |{\bf u}_i|^2,
\end{equation}
where ${\bf S}_i$ is the classical Heisenberg spin at the site $i$, ${\bf r}^0_{ij} \equiv {\bf r}^0_i-{\bf r}^0_j$, 
$c$ an elastic constant, $J_{\rm ex}$ the exchange interaction which is assumed to depend only on the distance between the two spins, and the summation $\langle i,j \rangle$ is taken over all the NN sites. As the displacement is usually small, i.e., $|{\bf u}_i|/|{\bf r}^0_i| \ll 1$, we can expand the exchange interaction with respect to the displacement as follows:
\begin{equation}\label{eq:expansion}
J_{\rm ex}\big(|{\bf r}^0_{ij} + {\bf u}_i-{\bf u}_j| \big) \simeq J_{\rm ex}\big(|{\bf r}^0_{ij}|\big) + \frac{d J_{\rm ex}}{dr}\Big|_{r=|{\bf r}^0_{ij}|} \, {\bf e}_{ij} \cdot ({\bf u}_i-{\bf u}_j ), 
\end{equation}
where ${\bf e}_{ij} \equiv {\bf r}^0_{ij}/|{\bf r}^0_{ij}|$ is the unit vector connecting NN sites $i$ and $j$. Substituting Eq. (\ref{eq:expansion}) into Eq. (\ref{eq:original_H}), we obtain
\begin{eqnarray}\label{eq:original_Hex}
{\cal H} &=& \sum_{\langle i,j \rangle }\Big[ J_{\rm ex}\big(|{\bf r}^0_{ij}|\big)+ \frac{d J_{\rm ex}}{dr}\Big|_{r=|{\bf r}^0_{ij}|} \, {\bf e}_{ij} \cdot ({\bf u}_i-{\bf u}_j ) \Big]{\bf S}_i \cdot {\bf S}_j \nonumber\\
& & + \frac{c}{2}\sum_i |{\bf u}_i|^2, \nonumber\\
&=&  \sum_{\langle i,j \rangle }J_{\rm ex}\big(|{\bf r}^0_{ij}|\big){\bf S}_i \cdot {\bf S}_j + \frac{c}{2}\sum_i |{\bf u}_i-{\bf u}^\ast_i|^2 - \frac{c}{2}\sum_i |{\bf u}^\ast_i|^2, \nonumber\\
{\bf u}^\ast_i &=& -\frac{1}{c}\sum_{j \in N(i) }\frac{d J_{\rm ex}}{dr}\Big|_{r=|{\bf r}^0_{ij}|} \, {\bf e}_{ij} \, \big( {\bf S}_i \cdot {\bf S}_j \big).
\end{eqnarray}
Here, $N(i)$ denotes all the NN sites of a site $i$. In Eq. (\ref{eq:original_Hex}), one can see the direct coupling between the lattice degrees of freedom ${\bf u}_i$ and the spin degrees of freedom ${\bf S}_i$. Integrating out the lattice degrees of freedom, or equivalently, minimizing the Hamiltonian with respect to ${\bf u}_i$, we can derive the effective spin interaction resulting from the SLC. Since the minimization condition is ${\bf u}_i={\bf u}^\ast_i$, the physical meaning of ${\bf u}^\ast_i$ is clear: it is the optimal local lattice distortion corresponding to the most probable ${\bf u}_i$-value. 

The NN sites on the breathing pyrochlore lattice contain two kinds of sites associated with small and large tetrahedra, and thus, we introduce two kinds of NN exchange interactions $J\equiv J_{\rm ex}\big(|{\bf r}^0_{ij}|_{\rm Small}\big)$ and $J'\equiv J_{\rm ex}\big(|{\bf r}^0_{ij}| _{\rm Large}\big)$. The degree of the breathing lattice-distortion is quantified by the ratio $J'/J \leq 1$. Then, the Hamiltonian reads ${\cal H} = {\cal H}_0 + {\cal H}_{\rm SL} $,
\begin{eqnarray}\label{eq:Hamiltonian}
{\cal H}_0 &=& J \, \sum_{\langle i,j \rangle_S } {\bf S}_i \cdot {\bf S}_j + J' \, \sum_{\langle i,j \rangle_L } {\bf S}_i \cdot {\bf S}_j, \nonumber\\
{\cal H}_{\rm SL} &=& -\frac{c}{2}\sum_i |{\bf u}^\ast_i|^2, \nonumber\\
{\bf u}^\ast_i &=& \Big\{ \sqrt{\frac{J \, b}{c}}\sum_{j \in N_S(i) } + \sqrt{\frac{J' \, b'}{c}}\sum_{j \in N_L(i) } \Big\} {\bf e}_{ij} \, \big( {\bf S}_i \cdot {\bf S}_j \big), \nonumber\\
\end{eqnarray}
where the NN sites $N_{S \, (L)}(i)$ and the summation $ \langle i,j \rangle _{S \, (L)}$ are defined only on the small (large) tetrahedra.  
The dimensionless parameters
\begin{eqnarray}\label{eq:b-def}
b &=& \frac{1}{cJ}\Big[ \frac{d J_{\rm ex}}{dr}\big|_{r=|{\bf r}^0_{ij}|_{\rm Small}} \Big]^2 \nonumber\\
b' &=& \frac{1}{cJ'}\Big[ \frac{d J_{\rm ex}}{dr}\big|_{r=|{\bf r}^0_{ij}|_{\rm Large}} \Big]^2 
\end{eqnarray}
measure the strength of the SLC for small and large tetrahedra, respectively. We take $J, \, J'>0$ and $d J_{\rm ex}/dr < 0$, so that $b, \, b' >0$. 
Note that the definition Eq. (\ref{eq:b-def}) indicates that $b'$ can be larger than $b$ depending on the $r$ dependence of $J_{\rm ex}(r)$.
In Eq. (\ref{eq:Hamiltonian}), ${\cal H}_0$ describes the NN exchange interactions, and the spin interaction mediated by the site phonon ${\cal H}_{\rm SL}$ can be rewritten more explicitly as
\begin{eqnarray}\label{eq:Hamiltonian_SL}
{\cal H}_{\rm SL} &=& - J \, b \, \sum_{\langle i,j \rangle_S } \big( {\bf S}_i \cdot {\bf S}_j \big)^2 - J' \, b' \,\sum_{\langle i,j \rangle_L } \big( {\bf S}_i \cdot {\bf S}_j \big)^2 \nonumber\\
&-& \sum_i \Big\{ \frac{Jb}{4} \sum_{j\neq k \in N_S(i)}+\frac{J'b'}{4}\sum_{j\neq k \in N_L(i)}\Big\} \big( {\bf S}_i \cdot {\bf S}_j \big)\big( {\bf S}_i \cdot {\bf S}_k \big) \nonumber\\
&-& \sqrt{J \, J' \,b \, b'}\sum_i \sum_{j \in N_S(i) } \sum_{k \in N_L(i) } {\bf e}_{ij} \cdot {\bf e}_{ik} \, \big( {\bf S}_i \cdot {\bf S}_j \big)\big( {\bf S}_i \cdot {\bf S}_k \big).  \nonumber\\
\end{eqnarray}
All the terms in ${\cal H}_{\rm SL}$ are quartic in ${\bf S}_i$. The first, second, and third lines describe the NN, intra-tetrahedra, and inter-tetrahedra interactions, respectively. The sign of the interactions in the first line is always negative, so that these $-({\bf S}_i \cdot {\bf S}_j)^2$ terms tend to align neighboring spins to be collinear and is known to be an origin of the spin nematic state. Spin collinearity can be measured by the quantity
\begin{equation}\label{eq:OP_nematic}
P = \frac{3}{2} \Big\langle \frac{1}{N^2}\sum_{i,j} \big( {\bf S}_i\cdot{\bf S}_j\big)^2 - \frac{1}{3} \Big\rangle, 
\end{equation}
where $N$ is a total number of spins and $\langle O \rangle$ denotes the thermal average of a physical quantity $O$. 
The spin and lattice-distortion correlations can be detected by measuring the spin structure factor
\begin{equation}\label{eq:F_S}
F_{\rm S}({\bf q}) = \Big\langle \Big| \frac{1}{N} \sum_i  {\bf S}_i \, e^{i{\bf q}\cdot{\bf r}^0_i}\Big|^2\Big\rangle 
\end{equation}
and the lattice-distortion structure factor
\begin{equation}\label{eq:F_L}
F_{\rm L}({\bf q}) = \Big\langle \Big| \frac{1}{N} \sum_i  {\bf u}^\ast_i \, e^{i{\bf q}\cdot{\bf r}^0_i}\Big|^2\Big\rangle .
\end{equation}
Since the breathing alternation has already been incorporated in the spin Hamiltonian, we will take ${\bf r}^0_i$ in Eqs. (\ref{eq:F_S}) and (\ref{eq:F_L}) as a regular position of the {\it uniform} pyrochlore lattice ignoring the bond-length alternation for simplicity. Because the local lattice distortion ${\bf u}^\ast_i$ and the spin ${\bf S}_i$ are related by Eq. (\ref{eq:Hamiltonian}), ${\bf u}^\ast_i$ is expected to order if the spins order, showing the associated Bragg peaks in $F_{\rm L}({\bf q})$.    

We note that the effective spin Hamiltonian in the bond-phonon model in which each bond length instead of the site position varies independently, is obtained in Eq. (\ref{eq:original_H}) by replacing $\sum_i |{\bf u}_i|^2$ with $\sum_{\langle i,j \rangle} |{\bf e}_{ij}\cdot\big({\bf u}_i-{\bf u}_j \big) |^2$ and by integrating out the bond-length degrees of freedom ${\bf e}_{ij}\cdot\big({\bf u}_i-{\bf u}_j \big)$. As a result, in the bond-phonon model, only the first line of Eq. (\ref{eq:Hamiltonian_SL}) is obtained. 

\section{effective Ising model}
\begin{figure}[t]
\includegraphics[scale=0.55]{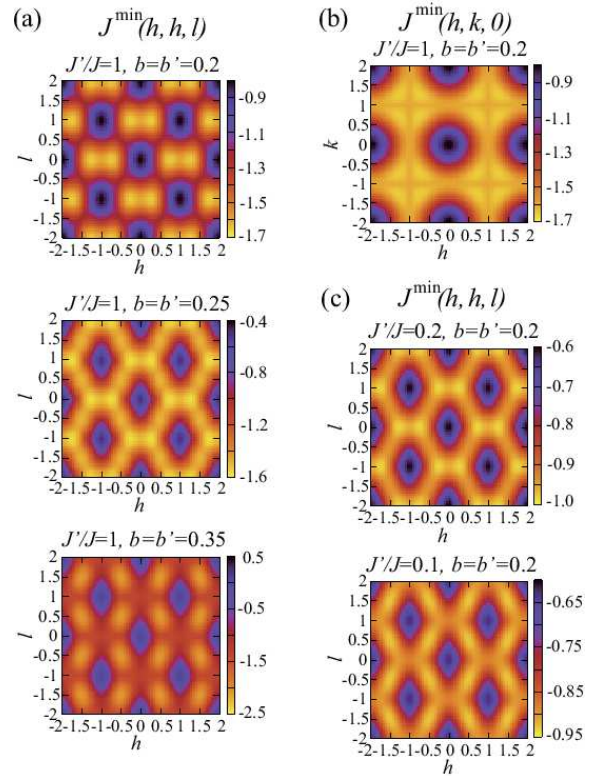}
\caption{Minimum eigen values of $J^{ab}({\bf q})$ in the $(h,h,l)$ [(a) and (c)] and $(h,k,0)$ [(b)] planes. In (a), $J'/J=1$ is fixed with varying $b=b'$. From top to bottom, $b=b'=0.2$, $0.25$, and $0.35$. In (b), $J'/J=1$ and $b=b'=0.2$. In (c), $b=b'=0.2$ is fixed with $J'/J=0.2$ (upper panel) and $0.1$ (lower one).  \label{fig:J_q}}
\end{figure}

\begin{figure}[t]
\includegraphics[scale=0.5]{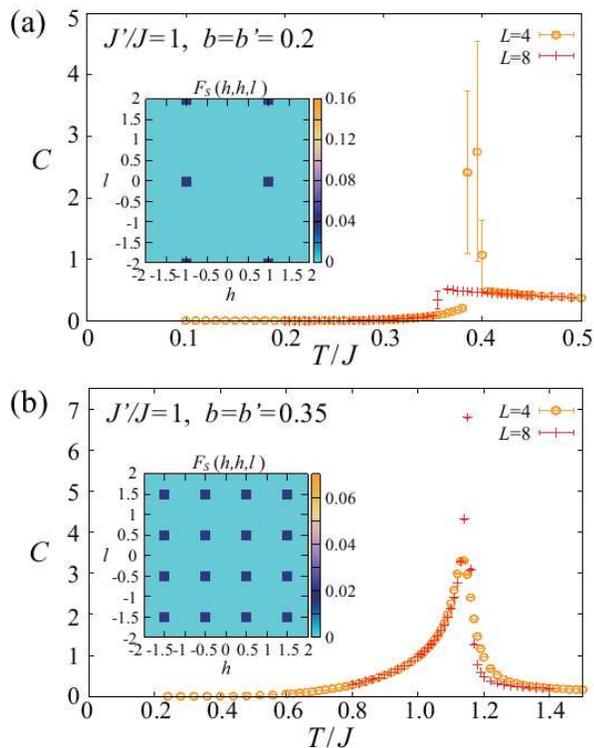}
\caption{MC results of the effective Ising model Eq. (\ref{eq:H_Ising}) with $J'/J=1$ obtained at $b=b'=0.2$ (a) and $b=b'=0.35$ (b). The main panel shows the temperature dependence of the specific heat $C$, and the inset of (a) [(b)] shows the spin structure factors $F_S({\bf q})$ in the $(h,h,l)$ plane obtained at $T/J=0.28$ ($T/J=0.8$) in the ordered state. \label{fig:Ising_MC}}
\end{figure}

In this section, we will discuss the ground-state ordering properties of the site-phonon Heisenberg model. To better understand the ordering behavior of the Heisenberg model, we introduce the effective Ising model and study its ordering properties. Namely, since the biquadratic terms with the negative sign $-({\bf S}_i\cdot {\bf S}_j)^2$ favor collinear spin states, which is actually verified by our MC simulations in the site-phonon Heisenberg model as will be shown in Secs. IV and V, we replace the Heisenberg spins ${\bf S}_i$ with the Ising variable $\sigma_i$, bearing collinear spin states in our mind. In the case without SLC, four spins on each tetrahedron satisfy the condition 
\begin{equation}\label{eq:2u2d}
\sigma_1+\sigma_2+\sigma_3+\sigma_4=0
\end{equation} 
to minimize the NN antiferromagnetic interaction, leading to a two-up and two-down spin configuration on each tetrahedron. In the case with SLC, our effective Ising spin Hamiltonian ${\cal H}_0+{\cal H}_{\rm SL}$ reads
\begin{eqnarray}\label{eq:H_Ising}
{\cal H}^{\rm eff}_{\rm Ising} &=& E_0 + J_{1}^{\rm eff} \sum_{\langle i,j \rangle_S} \sigma_i \sigma_j + J_{1'}^{\rm eff} \sum_{\langle i,j \rangle_L} \sigma_i \sigma_j \nonumber\\
&+& J_2^{\rm eff} \sum_{\langle \langle i,j \rangle \rangle} \sigma_i \sigma_j + J_3^{\rm eff} \sum_{\langle \langle \langle i,j \rangle \rangle \rangle} \sigma_i \sigma_j, \nonumber\\
E_0 &=& -N \, \frac{3}{2}(J b + J' b') \nonumber\\
J_{1}^{\rm eff} &=& J(1-b), \qquad J_{1'}^{\rm eff} = J'(1-b'), \nonumber\\
J_3^{\rm eff} &=& 2J_2^{\rm eff} = \sqrt{JJ' \, b \, b'},
\end{eqnarray} 
where $\langle \langle i,j \rangle \rangle$ and $\langle \langle \langle i,j \rangle \rangle \rangle$ respectively denote the summations over pairs of the second NN sites and the third NN ones along the bond directions. Thus, the ground-state spin correlation is determined by the $J_1$-$J_2$-$J_3$ Ising model. Due to the antiferromagnetic $J_3^{\rm eff}$, which is twice larger than $J_2^{\rm eff}$, three spins on a straight lines tend not to be parallel with each other \cite{Site_Bergman_06}. Note that for extremely strong SLC such that $b>1$ or $b'>1$, the NN interaction for small or large tetrahedra becomes ferromagnetic, although such a strong SLC might be unphysical. 

Now, we discuss the ordering properties of ${\cal H}^{\rm eff}_{\rm Ising}$. Introducing the Fourier transform $\sigma_i=\sum_{\bf q}\sigma^a_{\bf q}\exp(i{\bf q}\cdot {\bf r}_i)$ with the site index $i=(a, {\bf r}_i)$ and the sublattice index $a$, we can rewrite ${\cal H}^{\rm eff}_{\rm Ising} $ into the following form \cite{Reimers_MF, Okubo_pyro}
\begin{equation}\label{eq:H_Ising_FT}
{\cal H}^{\rm eff}_{\rm Ising}-E_0 = \frac{1}{2}\sum_{i,j} J_{ij} \sigma_i \sigma_j =  \frac{N}{8}\sum_{\bf q} \sum_{a,b=1}^4 J^{ab}({\bf q}) \sigma^a_{\bf q} \sigma^b_{-{\bf q}}. 
\end{equation}
Here, we have used the four-sublattice representation in which the sublattice indices $a=1, \, 2, \, 3, \, 4$ correspond to the four corners of a tetrahedron. Each component of $J^{ab}({\bf q})$ is given in Appendix A. Possible ordering wave-vectors can be obtained by analyzing eigen values of the 4-by-4 matrix $J^{ab}({\bf q})$, $J^{\rm min}({\bf q})$.
An ordering wave-vector with the lowest eigen value should be realized in the ground state, {\it provided that the local spin-length constraints $|\sigma_i|=1$ are satisfied}.

Figure \ref{fig:J_q} (a) shows $J^{\rm min}({\bf q})$ in the ${\bf q}=\frac{2\pi}{a}(h,h,l)$ plane in the uniform-pyrochlore case with $J'/J=1$ and $b=b'$. One can see that at $b'=b=0.2$, the eigen value is lowest near the wave vector $\pm(\frac{3}{4},\frac{3}{4},0)$ and that with increasing the strength of the SLC $b=b'$, the eigen value at the wave vector $(\frac{1}{2},\frac{1}{2},\pm\frac{1}{2})$, $J^{\rm min}(\frac{1}{2},\frac{1}{2},\frac{1}{2})$, gradually decreases, eventually leading to the minimum value at $b=b'=0.35$. We have checked that the lowest eigen value in the $(h,h,l)$ plane is the global minimum in the whole ${\bf q}$ space, which indicates that the system would try to take spin configurations characterized by these ordering wave-vectors. 

To see the validity of the above analysis respecting $J^{\rm min}({\bf q})$ but ignoring the local spin-length constraints, we perform MC simulations for the effective Ising model ${\cal H}^{\rm eff}_{\rm Ising}$ in the uniform-pyrochlore case with $J'/J=1$. 
Figure \ref{fig:Ising_MC} shows the MC results obtained for the effective Ising model ${\cal H}^{\rm eff}_{\rm Ising}$. At each temperature, we perform $5\times10^5$ Metropolis sweeps for observations after $5\times10^5$ sweeps for thermalization under periodic boundary conditions and the statistical average is taken over $4$ independent runs. One can see from the insets of Figs. \ref{fig:Ising_MC}(a) and (b) that the low-temperature ordered phase at the weak SLC with $b=b'=0.2$ is characterized by the $(1,1,0)$-type Bragg peaks, while the one at the strong SLC with $b=b'=0.35$ is characterized by the $(\frac{1}{2},\frac{1}{2},\frac{1}{2})$-type one, suggesting that the analysis based on $J^{\rm min}({\bf q})$ is valid only in the strong SLC regime where the resulting $(\frac{1}{2},\frac{1}{2},\frac{1}{2})$-type order satisfies the spin-length condition. By contrast, the $(\frac{3}{4},\frac{3}{4},0)$-type order expected from the $J^{\rm min}({\bf q})$ analysis does not satisfy the spin-length condition and cannot be stabilized. Note that as we will see in the following section, this situation is also the case in the original site-phonon Heisenberg model. Namely, the $(\frac{1}{2},\frac{1}{2},\frac{1}{2})$ order is realized at stronger SLC, while the $(1,1,0)$ order is realized instead of the $(\frac{3}{4},\frac{3}{4},0)$ one at weaker SLC. The obtained $(1,1,0)$ order is favored probably due to the order-by-disorder mechanism, because in $J^{\rm min}({\bf q})$ in the ${\bf q}=\frac{2\pi}{a}(h,k,0)$ plane shown in Fig. \ref{fig:J_q}(b), the line degeneracy arises along $(1,k,0)$.   

Next, we shall discuss the effect of the breathing alternation $J'/J$ on the ordering properties. Figure \ref{fig:J_q}(c) shows $J^{\rm min}(h,h,l)$ for the breathing pyrochlore lattice with $J'/J=0.2$ and $0.1$ at the weak SLC $b'=b=0.2$. Compared with the corresponding uniform case with $J'/J=1$ shown in the upper panel of Fig. \ref{fig:J_q}(a), the eigen values $J^{\rm min}(\frac{1}{2},\frac{1}{2},\frac{1}{2})$ and $J^{\rm min}(1,1,0)$ are competing with each other in the breating cases, and at $J'/J=0.1$, $J^{\rm min}(\frac{1}{2},\frac{1}{2},\frac{1}{2})$ becomes lower than $J^{\rm min}(1,1,0)$, suggesting that the $(\frac{1}{2},\frac{1}{2},\frac{1}{2})$ state can easily show up in the strongly breathing case. Since ordering vectors other than the $(1,1,0)$ and $(\frac{1}{2},\frac{1}{2},\frac{1}{2})$ families do not come into play, we will examine the relative stability between these two states.   

For the uniform pyrochlore lattice with $J'/J=1$, $J^{\rm min}(1,1,0)=-2J_1^{\rm eff}=-2J(1-b)$ and $J^{\rm min}(\frac{1}{2},\frac{1}{2},\frac{1}{2})=-6J_3^{\rm eff}=-6Jb$, so that the phase boundary between the $(1,1,0)$ and $(\frac{1}{2},\frac{1}{2},\frac{1}{2})$ states is given by $b=1/4$: for $b<1/4$ the $(1,1,0)$ state is stable, while for $b>1/4$, the $(\frac{1}{2},\frac{1}{2},\frac{1}{2})$ state is realized. This simple analysis is consistent with the MC result obtained in the original site-phonon Heisenberg model \cite{Site_AK_16}.

We shall turn to the breathing pyrochlore lattice with $J'/J \leq 1$. The concrete expression of the lowest eigen value is given in Appendix A. To see the $J'/J$-dependence, we first consider the simplified case of $b'=b$ ($b, \, b' <1$). Comparing $J^{\rm min}(1,1,0)$ with $J^{\rm min}(\frac{1}{2},\frac{1}{2},\frac{1}{2})$, we obtain the following condition of the $(1,1,0)$ spin state being more stable:
\begin{eqnarray}\label{eq:boundary}
\sqrt{\frac{J'}{J}} &\geq& r-\sqrt{r^2-1} , \quad r = \frac{3}{8}\frac{1-2b}{b(1-b)}, \nonumber\\
b &\leq & \frac{1}{4} .
\end{eqnarray}
This phase boundary in the $b-J'/J$ plane is depicted by an orange curve in Fig. \ref{fig:T0diagram}(a). Noting that $r-\sqrt{r^2-1}$ depends only on $b$, one finds that for a fixed value of $b$, the $(1,1,0)$ state becomes unstable against the $(\frac{1}{2},\frac{1}{2},\frac{1}{2})$ state with decreasing the ratio $J'/J$. Furthermore, since $r-\sqrt{r^2-1} \rightarrow 0$ in the $b\rightarrow 0$ limit, the phase boundary always exists as long as $b\neq 0$. Thus, in principle, even for weak SLC, the $(\frac{1}{2},\frac{1}{2},\frac{1}{2})$ state becomes favorable if $J'/J$ can be tuned to a sufficiently small value. Of course, in reality, the SLC parameters $b$ and $b'$ are correlated to $J'$ and $J$ in the form of Eq. (\ref{eq:b-def}), so that independent parameter tuning would be difficult.      

In the general case of $b\neq b'$ ($b, \, b'<1$), the stability condition of the $(1,1,0)$ state is given by
\begin{eqnarray}\label{eq:boundary_general}
&& R_- \leq \sqrt{\frac{J'}{J}} \leq \min(1, R_+), \nonumber\\
&& R_\pm = \tilde{r} \pm \sqrt{ \tilde{r}^2 - \frac{1-b}{1-b'}  }, \quad \tilde{r} = \frac{3}{8}\Big( \frac{1-b}{\sqrt{bb'}}-\frac{\sqrt{bb'}}{1-b'} \Big). \nonumber\\
\end{eqnarray}
The phase boundary (\ref{eq:boundary_general}) in the $b-b'-J'/J$ space is presented by orange mesh in Fig. \ref{fig:T0diagram}(b). One can see that this simple analytic result for the {\it effective Ising model} is in good agreement with the numerical results for the {\it original site-phonon Heisenberg model} depicted by colored points in Fig. \ref{fig:T0diagram}(b).  


\section{Monte Carlo Result in the Weak SLC Regime}
\begin{figure}[t]
\includegraphics[width=\columnwidth]{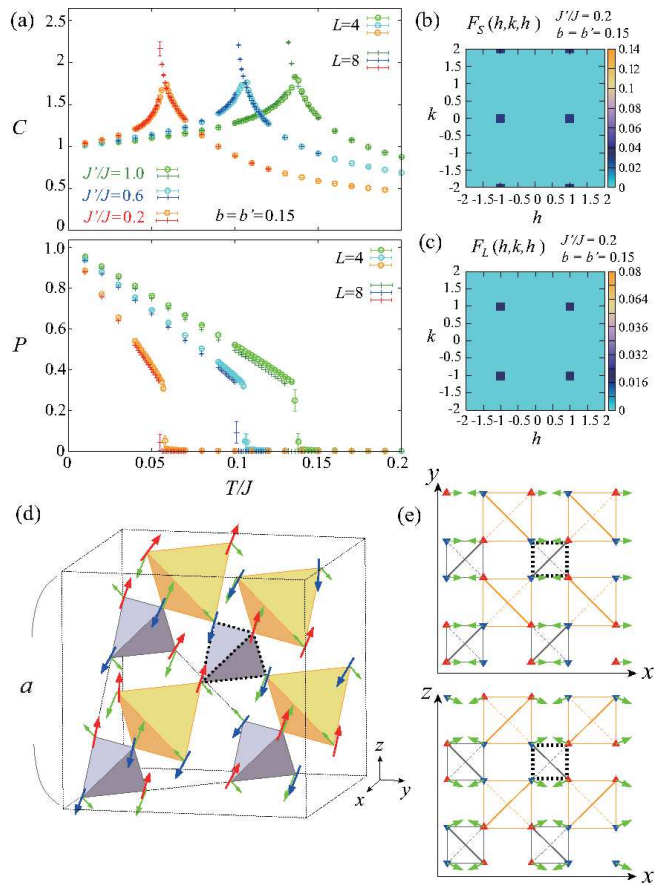}
\caption{MC results obtained in the weak SLC regime at $b=b'=0.15$. (a) The temperature dependences of the specific heat $C$ (upper panel) and the spin collinearity $P$ (lower panel). Green, blue, and red colored symbols denote the data for $J'/J=1$, $0.6$, and $0.2$, respectively. (b) and (c) Spin and lattice-distortion structure factors $F_S({\bf q})$ and $F_L({\bf q})$ in the ordered state in the $(h,k,h)$ plane at $T/J=0.03$ for $J'/J=0.2$ and $L=8$. (d) Corresponding spin and lattice-distortion snapshots at $T/J=0.01$. Red (blue) arrows represent up (down) spins and green arrows represent local lattice distortions ${\bf u}^\ast_i$. A thin dotted box denotes a cubic unit cell of edge length $a$. (e) A snapshot of ${\bf u}^\ast_i$ on the $xy$- and $xz$-planes of the lattice, where green arrows in the upper and lower panels represent the $xy$- and $xz$-components of ${\bf u}^\ast_i$, respectively. Gray (orange) colored boxes correspond to gray (yellow) colored tetrahedra in (d), and red (blue) triangles denote the up (down) spins in (d). The small tetrahedron at the center of the cubic unit cell is outlined by black dots in (d) and (e). \label{fig:weak}}
\end{figure}

In the previous section, we discussed the ordering properties of collinear spin states by analyzing the effective Ising model. Now, we shall return to the original site-phonon Heisenberg model and investigate its low temperature magnetic properties by means of MC simulations. In our MC simulations, we basically perform $10^6$ Metropolis sweeps at each temperature in the cooling run under periodic boundary conditions, where the first half is discarded for thermalization. 
In a single spin flip at each site, we first try to update a spin to a randomly chosen new spin state, and subsequently try to rotate a spin by the angle $\pi$ around the local mean field. These two updates are performed by the standard Metropolis algorithm \cite{Loop_Shinaoka_14}.
Observations are done in every $5$ MC steps and the statistical average is taken over $8$ independent runs. Since the cubic unit cell contains $16$ sites [see Fig. \ref{fig:weak}(d)], a total number of spins $N$ is $N=16 L^3$ for a system size $L$.

By measuring various physical quantities such as the spin collinearity $P$ and the spin and lattice-distortion structure factors $F_{\rm S}({\bf q})$ and $F_{\rm L}({\bf q})$, we identified the low-temperature ordered phase for various sets of the parameters $b$, $b'$, and $J'/J$. We find four different types of spin-lattice-coupled orders, the stability regions of which are summarized in Fig. \ref{fig:T0diagram}. In the smaller $b$ and $b'$ (weaker SLC) region, the tetragonal-symmetric long-range-order characterized by $(1,1,0)$ magnetic Bragg peaks are realized, while in the relatively larger $b$ and $b'$ (stronger SLC) region, we found three different types of ordering patterns, all of which are characterized by $(\frac{1}{2},\frac{1}{2},\frac{1}{2})$ magnetic Bragg reflections. In this section, we will discuss the $(1,1,0)$ state realized in the weaker SLC regime.

In the weak SLC regime, the ordering properties of the uniform pyrochlore lattice are not altered by the breathing lattice-distortions. Figure \ref{fig:weak}(a) shows temperature dependences of the specific heat $C$ and the spin collinearity $P$ for $b=b'=0.15$ in the weak SLC regime. One can see that the system undergoes a first-order transition into a collinear spin state in all cases. Although the transition temperature decreases with decreasing $J'/J$, the first-order character remains unchanged.
The spin structures in the low-temperature ordered phases are characterized by the $(1,1,0)$-type Bragg peaks. Figure \ref{fig:weak}(b) shows a typical spin structure factor $F_{\rm S}({\bf q})$ obtained in the breathing case of $J'/J=0.2$, where high-intensity peaks are found in the $(h,k,h)$ plane, i.e., the magnetic Bragg peaks at $\pm(1,0,1)$ and $\pm(1,2,1)$. If the spin order is cubic symmetric, one should find same-height Bragg peaks at all the cubic-symmetric points $\pm(1,1,0)$, $\pm(1,0,1)$, and $\pm(0,1,1)$. The obtained ordered phase, however, picks up only one of the three, namely, $\pm(1,0,1)$, so that its magnetic structure is tetragonal symmetric. This argument can be confirmed also from the corresponding real-space spin configuration shown in Fig. \ref{fig:weak}(d). The $\uparrow\downarrow\uparrow\downarrow$ chains are running along the facing two tetrahedral bonds, i.e., the $[101]$ and $[\overline{1}01]$ directions, while the $\uparrow\uparrow\downarrow\downarrow$ chains are running along the remaining four tetrahedral bonds ,i.e., the $[110]$, $[1\overline{1}0]$, $[011]$, and $[01\overline{1}]$ directions. 

The associated local-lattice-distortion pattern also reflects this tetragonal symmetry. As shown in Fig. \ref{fig:weak}(e), the local lattice-distortion vector ${\bf u}_i^\ast$ does not have the $y$-component and is restricted only in the $xz$-plane showing the commensurate behavior which is reflected in the lattice-distortion structure factor $F_{\rm L}({\bf q})$ as the Bragg peaks at all $(\pm 1,\pm 1,\pm 1)$ points [Fig. \ref{fig:weak}(c)]. 

Although the $y$ direction is special in this particular spin configuration, which one is selected among the three equivalent points $\pm(1,1,0)$, $\pm(1,0,1)$, and $\pm(0,1,1)$ depends on the initial condition of the MC simulation. Thus, we call the ordered state ``the $(1,1,0)$ state'' independent of selected directions. We note that in our MC simulations, we often encounter domain states consisting of two kinds of different $(1,1,0)$ domains such as $(1,1,0)$ and $(0,1,1)$ domains \cite{Site_AK_16}. 

Now, we look into the real-space spin configuration in units of tetrahedron. The collinear $(1,1,0)$ state consists only of two-up and two-down (2u2d) tetrahedra. This is because weaker SLC yields weaker effective further neighbor interactions and thus the spin ordering is strongly subject to the local constraint due to the NN interaction, Eq. (\ref{eq:2u2d}). Concerning the local lattice distortion, all the small and large tetrahedra respectively have the same distortions [see green arrows in Figs. \ref{fig:weak}(d) and (e)]. Such a situation is also true for the $(1,1,0)$ state realized on the {\it uniform} pyrochlore lattice \cite{Site_AK_16}. 

From this classification of the lattice distortion, the relation between the present site-phonon model and the phenomenological theory of SLC in the uniform pyrochlore antiferromagnets \cite{SLC_Tchernyshyov_prb_02} becomes clear. The basic assumption of the latter approach is that all tetrahedra of the same orientation have the same lattice distortion. As the lattice distortion in the $(1,1,0)$ state satisfies this assumption, the site-phonon model is equivalent to the phenomenological theory in which the $(1,1,0)$ state has been discussed \cite{SLC_Tchernyshyov_prl_02, SLC_Tchernyshyov_prb_02, SLC_Chern_06}. Note that the corresponding phenomenological theory for the breathing pyrochlore lattice is not available, and even on the {\it uniform} pyrochlore lattice, this equivalence holds only for the weak SLC. Stronger SLC, on the other hand, induces stronger effective further neighbor interactions, so that in the strong SLC regime, the local constraint Eq. (\ref{eq:2u2d}) is not satisfied any more, leading to the $(\frac{1}{2},\frac{1}{2},\frac{1}{2})$ magnetic orders consisting of various types of tetrahedral configurations. In the next section, we will discuss the $(\frac{1}{2},\frac{1}{2},\frac{1}{2})$ state realized in the strong SLC regime, which is not accessible by the phenomenological theory even for the uniform pyrochlore lattice.     


\section{Monte Carlo Result in the Strong SLC Regime}
The analytical result in Sec. III indicates that the $(\frac{1}{2},\frac{1}{2},\frac{1}{2})$ magnetic order is favored with increasing the strength of either the SLC (increasing $b$ and $b'$) or the breathing alternation (decreasing $J'/J$). This is supported by our MC simulations of the site-phonon Heisenberg model. As summarized in Fig. \ref{fig:T0diagram}, the $(\frac{1}{2},\frac{1}{2},\frac{1}{2})$-type magnetic Bragg peaks are observed basically at stronger SLC. When the breathing lattice-distortion is strong enough such that $J'/J=0.1$, the Bragg peaks of this type are observed also at moderate SLC.  
The $(\frac{1}{2},\frac{1}{2},\frac{1}{2})$-type magnetic order has three distinct types, which we call type I, II, and III as shown in Fig. \ref{fig:T0diagram}.   
It will be shown that in all the cases, spins order collinearly through a first-order transition and that the three types of the $(\frac{1}{2},\frac{1}{2},\frac{1}{2})$ states have different tetrahedral configurations (see Table.\ref{table:tetra_conf}). In particular, the type-II and type-III orders are induced by the breathing lattice-alternation, and the massive ground-state degeneracy inherent to their non-cubic spin structures might show up as the macroscopic residual entropy. 

\subsection{The type-I $(\frac{1}{2},\frac{1}{2},\frac{1}{2})$ magnetic order}
\begin{figure}[t]
\includegraphics[width=\columnwidth]{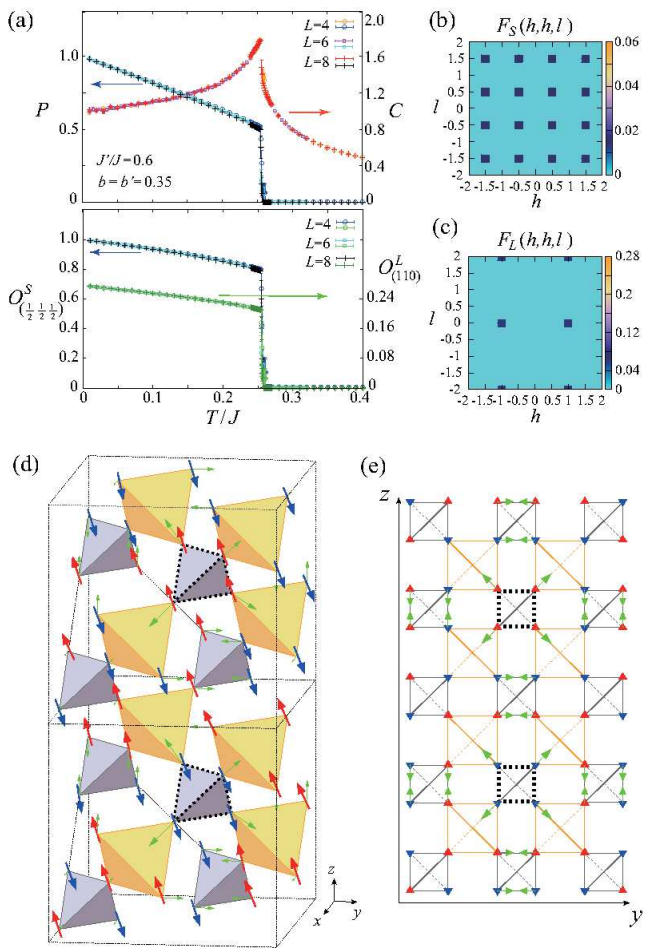}
\caption{MC results for the type-I order obtained in the strong SLC regime at $b=b'=0.35$ and $J'/J=0.6$. (a) The temperature dependences of the specific heat $C$, the spin collinearity $P$ (upper panel), and the average spin and lattice Bragg intensities $O^S_{(\frac{1}{2}\frac{1}{2}\frac{1}{2})}$ and $O^L_{(110)}$  (lower one). (b) and (c) Spin and lattice-distortion structure factors $F_S({\bf q})$ and $F_L({\bf q})$ in the ordered state in the $(h,h,l)$ plane at $T/J=0.15$ for $L=8$. (d) and (e) Spin and lattice-distortion snapshots at $T/J=0.01$, where in (e), the $yz$-component of ${\bf u}^\ast_i$ on the $yz$ plane of the lattice is shown. The small tetrahedron at the center of the cubic unit cell is outlined by black dots in (d) and (e). Notations are the same as those in Fig. \ref{fig:weak}.  \label{fig:typeI}}
\end{figure}

The type-I order is realized in the wide range of the parameter space including $J'/J=1$ which corresponds to the uniform pyrochlore lattice. As an example of the type-I order, the MC result for $b=b'=0.35$ and $J'/J=0.6$ is shown in Fig. \ref{fig:typeI}. One can see from the upper panel of Fig. \ref{fig:typeI} (a) that the system exhibits a first-order transition into a collinear spin state with $P \neq 0$, and from Fig. \ref{fig:typeI} (b) that the low-temperature phase is characterized by the magnetic Bragg peaks appearing at $(\frac{1}{2} + n_h,\frac{1}{2}+n_h,\frac{1}{2}+n_l )$ with $n_h$ and $n_l$ integers. We have checked that the spin structure factor $F_{\rm S}({\bf q})$ has the Bragg peaks of the same height at all the cubic-symmetric families of $(\frac{1}{2},\frac{1}{2},\frac{1}{2})$, i.e., $(\pm\frac{1}{2},\pm\frac{1}{2},\pm\frac{1}{2})$, and thus, the ordered phase is a cubic-symmetric multiple-${\bf q}$ state. 
The real-space spin configuration is shown in Fig. \ref{fig:typeI}(d). The type-I order is constructed by the alternating array of the two types of cubic unit cells, i.e., the upper and lower cubic unit cells in Fig. \ref{fig:typeI}(d). Thus, the type-I $(\frac{1}{2},\frac{1}{2},\frac{1}{2})$ magnetic order is composed of $\uparrow\uparrow\downarrow\downarrow$ spin-chains running along all the tetrahedral bonds, namely, the $[110]$ and cubic-symmetric directions \cite{Site_AK_16}.
Once the spin configuration is fixed, the local lattice distortions ${\bf u}^\ast_i$ is given by Eq. (\ref{eq:Hamiltonian}). The associated local lattice distortion is also cubic-symmetric, as ${\bf u}^\ast_i$ projected on the $xy$ and $xz$ planes look the same as the $yz$ projection shown in Figs. \ref{fig:typeI}(e).  The commensurate periodic pattern of ${\bf u}^\ast_i$ is signaled by the multiple Bragg peaks of equal heights at all $(\pm 1,\pm 1,0)$, $(\pm 1,0,\pm 1)$, and $(0,\pm 1, \pm 1)$ points in the lattice-distortion structure factor $F_{\rm L}({\bf q})$ [see Fig. \ref{fig:typeI}(c)]. In the lower panel of Fig. \ref{fig:typeI} (a), we show the temperature dependences of the averaged values of the Bragg peaks for $F_{\rm S}({\bf q})$ and $F_{\rm L}({\bf q})$, which are respectively defined by $O^{\rm S}_{(\frac{1}{2}\frac{1}{2}\frac{1}{2})}=2\sum_{h,k,l=\pm1/2}F_{\rm S}(h,k,l)$ and $O^{\rm L}_{(110)}=(1/12)\sum_{h,k=\pm1}\big[ F_{\rm L}(h,k,0)+F_{\rm L}(h,0,k)+F_{\rm L}(0,h,k) \big]$. Both $O^{\rm S}_{(\frac{1}{2}\frac{1}{2}\frac{1}{2})}$ and $O^{\rm L}_{(110)}$ evolve with decreasing the temperature below the first-order transition temperature, so that these two quantities serve as the order parameters of this spin-lattice-coupled order.  

In units of tetrahedron, this spin state is composed of 6 two-up and two-down (2u2d), 1 four-up (4u), and 1 four-down (4d) small tetrahedra and 4 three-up and one-down (3u1d) and 4 one-up and three-down (1u3d) large tetrahedra  (see Table \ref{table:tetra_conf}). As one can see from Fig. \ref{fig:typeI}(d), one cubic unit cell and a neighboring one respectively contain 4u and 4d tetrahedra at each center [see the small tetrahedra outlined by black dots in Fig. \ref{fig:typeI}(d)], which are surrounded by down and up spins, respectively. Small tetrahedra other than the 4u and 4d ones are the 2u2d one. As we will see below, the spin configurations of the central small tetrahedra differ among the type-I, type-II, and type-III orders.

\subsection{The type-II and type-III $(\frac{1}{2},\frac{1}{2},\frac{1}{2})$ magnetic orders}
\begin{figure}[t]
\includegraphics[width=\columnwidth]{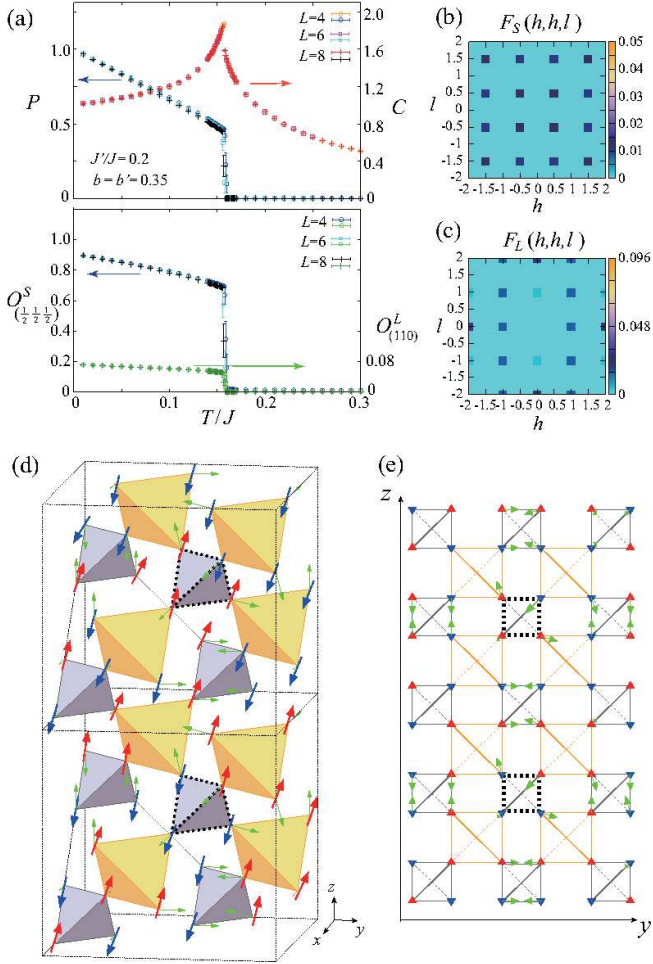}
\caption{MC results for the type-II order obtained in the strong SLC regime at $b=b'=0.35$ and $J'/J=0.2$. Notations are the same as those in Fig. \ref{fig:typeI}. The structure factors [(b) and (c)] and the snapshots [(d) and (e)] are obtained at $T/J=0.1$ and at $T/J=0.01$, respectively.  \label{fig:typeII}}
\end{figure}

\begin{figure}[t]
\includegraphics[width=\columnwidth]{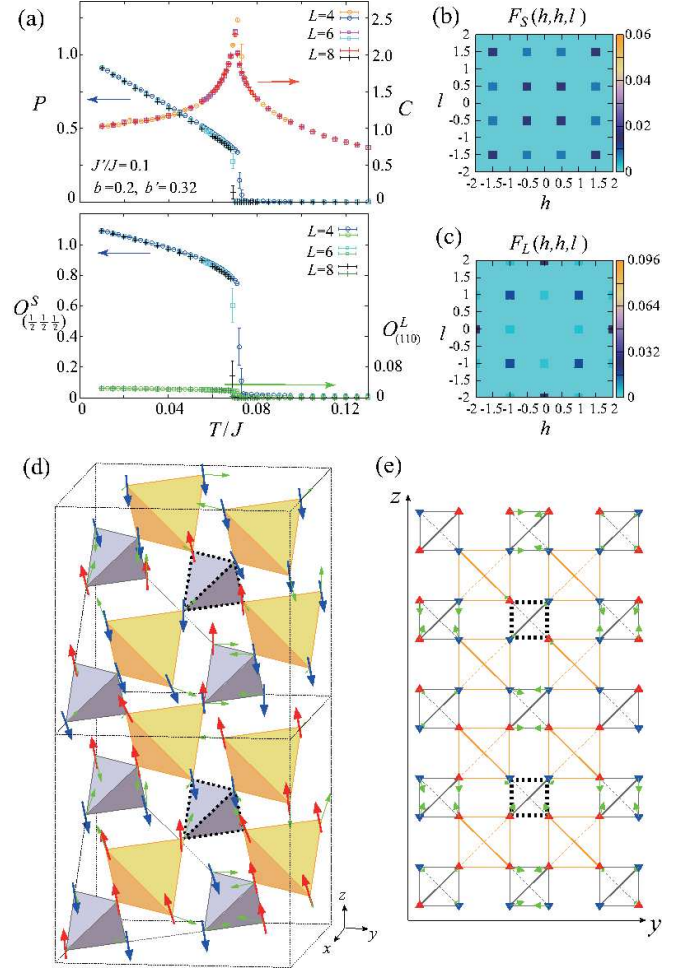}
\caption{MC results for the type-III order obtained in the strong SLC regime at $b=0.2$, $b'=0.32$, and $J'/J=0.1$. Notations are the same as those in Fig. \ref{fig:typeI}. The structure factors [(b) and (c)] and the snapshots [(d) and (e)] are obtained at $T/J=0.06$ and at $T/J=0.01$, respectively. \label{fig:typeIII}}
\end{figure}

The type-II and type-III orders are realized on the breathing pyrochlore lattice with strong bond-alternations. As one can see from Fig. \ref{fig:T0diagram}, the type-II order is realized in the strong-SLC case with small values of $J'/J$, whereas the type-III order is realized in the moderate-SLC case with much smaller values of $J'/J$, e.g., $J'/J=0.1$. In particular, in the general case of $b\neq b'$, the type-III order can be stabilized for relatively small $b$ if $b'$ is large enough, as is demonstrated in Fig. \ref{fig:typeIII}. 
Note that, in contrast to the type-I order, the type-II and type-III orders are realized only in the breathing-pyrochlore case. Typical examples of these two orders are shown in Figs. \ref{fig:typeII} and \ref{fig:typeIII}. In both cases, a first-order transition into a collinear spin state occurs, and the low-temperature ordered phase is characterized by the $(\frac{1}{2},\frac{1}{2},\frac{1}{2})$-type magnetic Bragg peaks in $F_{\rm S}({\bf q})$. The peak heights at $(\pm\frac{1}{2},\pm\frac{1}{2},\pm\frac{1}{2})$ and $(\pm\frac{3}{2},\pm\frac{3}{2},\pm\frac{3}{2})$ are all the same, but are slightly higher than those at $(\pm\frac{3}{2},\pm\frac{3}{2},\pm\frac{1}{2})$ and $(\pm\frac{1}{2},\pm\frac{1}{2},\pm\frac{3}{2})$ families, in contrast to the type-I case where all the $(\frac{1}{2}+n_h,\frac{1}{2}+n_k,\frac{1}{2}+n_l)$ ($n_h,n_k, n_l:{\rm integer}$) are equivalent. This is reflected in the temperature dependence of the averaged value of the $(\frac{1}{2},\frac{1}{2},\frac{1}{2})$ magnetic Bragg peaks. As one can see from the lower panels of Figs. \ref{fig:typeII}(a) and \ref{fig:typeIII}(a), $O^S_{(\frac{1}{2}\frac{1}{2}\frac{1}{2})}$ in the type-II and type-III orders approaches a value slightly deviated from unity, implying that the spin ordering patterns differ from the type-I structure. The change in the ordering pattern can clearly be seen in the associated lattice distortion. In Figs. \ref{fig:typeII} and \ref{fig:typeIII}, the lattice-distortion order parameter for the type-I order $O^L_{(110)}$ is strongly suppressed in the type-II and type-III orders, and instead, the $(1,1,1)$-type Bragg peaks show up in $F_{\rm L}({\bf q})$ as can be seen from Figs. \ref{fig:typeII}(c) and \ref{fig:typeIII}(c). 

The MC snapshots taken in the low-temperature ordered phases are shown in Figs. \ref{fig:typeII}(d) and \ref{fig:typeIII}(d). Although it is difficult to figure out ordering rules for spin itself, all the three phases characterized by the $(\frac{1}{2},\frac{1}{2},\frac{1}{2})$ magnetic Bragg peaks obey a common rule in units of tetrahedron, an alternating array of the two kinds of the cubic unit cells. As mentioned above, in the type-I order, each cubic unit cell of one kind has the 4u small tetrahedron at its center, whereas each of the other kind has the 4d one. In the type-II order, the 4u and 4d central tetrahedra are replaced with the 3u1d and 1u3d ones, respectively, while the 2u2d off-center small tetrahedra remain unchanged. In the type-III order, both the 4u and 4d central tetrahedra are replaced with the 2u2d one keeping the off-center small tetrahedra the same, and thus, all the small tetrahedra are the 2u2d one. The ratio of the numbers of the 4u, 3u1d, 2u2d, 1u3d, and 4d tetrahedra is summarized in Table \ref{table:tetra_conf}. Because of the configuration change in the central small tetrahedra from the isotropic 4u (4d) into the anisotropic 3u1d (1u3d) or 2u2d, spin chains along the $[110]$ and its cubic-symmetric directions become non-equivalent, namely, the spin states in the type-II and type-III orders become non-cubic. This symmetry lowering is reflected in the associated local lattice distortions. As shown in Figs. \ref{fig:typeII}(e) and \ref{fig:typeIII}(e), ${\bf u}^\ast_i$ on the central small tetrahedra are strongly disturbed and the cubic symmetry of ${\bf u}^\ast_i$ is not kept any more.

\begin{table}[b]
\caption{Tetrahedral spin configurations, the total number of states (NOS), and the resultant residual entropy per spin $s^R$ in the type-I, type-II, and type-III orders. \label{table:tetra_conf}}

\begin{tabular}{|c|c| c c c c c|c|c|}
\hline
type & tetra. size & 4u & 3u1d & 2u2d & 1u3d & 4d & NOS & $s^R/k_{\rm B}$\\
\hline
\multirow{2}{*}{I} & small & 1 : & 0 : & 6 : & 0 : & 1  & \multirow{2}{*}{${\cal O}(1)$} & \multirow{2}{*}{$0$}    \\ 
 & large & 0 : & 4 : & 0 : & 4 : & 0   &   &   \\
\hline
\multirow{2}{*}{II} & small & 0 : & 1 : & 6 : & 1 : & 0 & \multirow{2}{*}{${\cal O}(1)\times 4^{L^3}$} & \multirow{2}{*}{$\frac{\ln (4)}{16}$}    \\ 
 & large & 1 : & 3 : & 0 : & 3 : & 1   &   &       \\
\hline
\multirow{2}{*}{III} & small & 0 : & 0 : & 8 : & 0 : & 0 & \multirow{2}{*}{${\cal O}(1)\times 6^{L^3}$} & \multirow{2}{*}{$\frac{\ln (6)}{16}$}    \\ 
 & large & 2 : & 2 : & 0 : & 2 : & 2    &   &        \\
\hline
\end{tabular}
\end{table} 

Now that the ordering patterns of the three $(\frac{1}{2},\frac{1}{2},\frac{1}{2})$ states are unveiled, we shall discuss why the type-II and type-III structures are induced by the breathing lattice-distortions. Bearing the ground states of the three types of collinear magnetic orders in our mind, we will examine relative stabilities among them based on the effective Ising model, Eq. (\ref{eq:H_Ising}). With increasing the breathing lattice-distortion, namely, decreasing $J'/J$, the ratio $J^{\rm eff}_{1}/J^{\rm eff}_{1'} = \frac{J}{J'}\frac{1-b}{1-b'}$ increases, so that the NN antiferromagnetic interaction for small tetrahedra $J^{\rm eff}_{1}$ tends to be more respected and all the small tetrahedra try to satisfy the local constraint Eq. (\ref{eq:2u2d}). As a result, with decreasing $J'/J$, the 4u (4d) central small tetrahedra in the type-I order become 3u1d (1u3d) and ultimately 2u2d, which respectively correspond to the type-II and type-III orders. At the level of a single tetrahedron, there are 1, 4, and 6 possible spin configurations for the 4u or 4d, the 3u1d or 1u3d, and the 2u2d tetrahedra, respectively. As we will see below, in the type-II and type-III orders, the 4-fold and 6-fold degeneracies at each central small tetrahedron result in the macroscopic residual entropy of the whole system.

\subsection{Residual entropy}
\begin{figure}[t]
\includegraphics[scale=0.50]{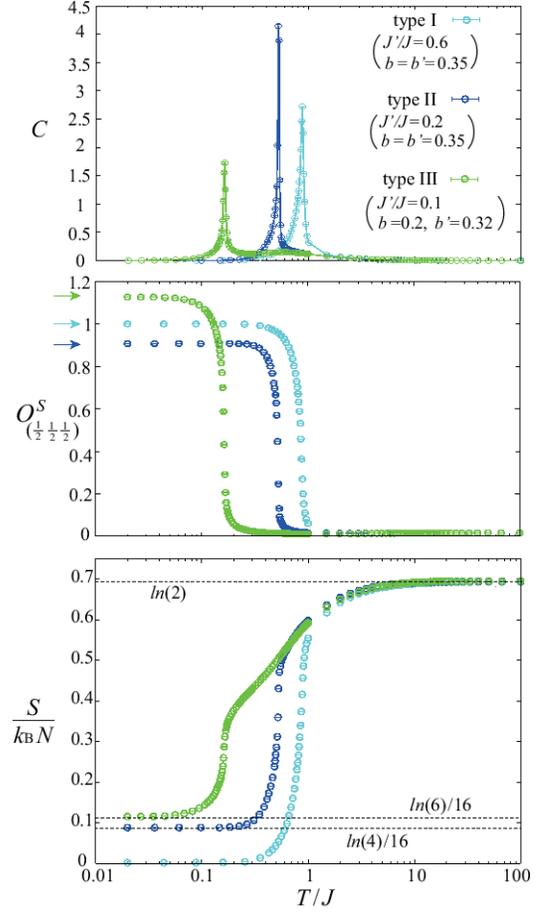}
\caption{The temperature dependences of the specific heat $C$ (upper panel), the average magnetic Bragg intensity $O^S_{(\frac{1}{2}\frac{1}{2}\frac{1}{2})}$ (middle panel), and the entropy per spin $S/(k_{\rm B}N)$ (lower panel) obtained in the effective Iing model Eq. (\ref{eq:H_Ising}) with $L=4$. Cyan, blue, and green points denote the data for the type-I ($b=b'=0.35$ and $J'/J=0.6$), type-II ($b=b'=0.35$ and $J'/J=0.2$), and type-III ($b=0.2$, $b'=0.32$, and $J'/J=0.1$) orders, respectively. Error bars are smaller than symbols. In the middle panel, horizontal arrows represent the $T=0$ $O^S_{(\frac{1}{2}\frac{1}{2}\frac{1}{2})}$-values of the corresponding site-phonon Heisenberg model. In the lower panel, dashed lines represent $S/(k_{\rm B}N) = \ln(2)$ (top), $\frac{\ln(6)}{16}$ (middle), and $\frac{\ln(4)}{16}$ (bottom).  \label{fig:entropy}}
\end{figure}

As mentioned above, the difference among the type-I, type-II, and type-III orders consists in the spin configuration at the central small tetrahedra in the cubic unit cell. In their ground states where the collinearly aligned spins could be regarded as Ising spins, the number of states (NOS) of the {\it isolated} 3u1d or 1u3d central small tetrahedron is 4 and that of the 2u2d one is 6, while the NOS of the 4u or 4d one is 1. In the type-I order, as the total NOS is ${\cal O}(1)$, the entropy per spin vanishes in the thermodynamic limit. The question is how many NOS the ground state has if the 4u (4d) central small tetrahedra in the type-I order is replaced with the 3u1d (1u3d) or 2u2d one which respectively correspond to the type-II and type-III orders. In the original type-I order shown in Fig. \ref{fig:typeI}(d), the 4u (4d) tetrahedron is surrounded by down (up) spins and all the surrounding spins are connected to the outer 2u2d tetrahedra, constituting $\uparrow\uparrow\downarrow\downarrow$ chains. Thus, the four spins on the 4u or 4d tetrahedron have the same numbers of the second-NN up and down spins and the third-NN ones. Since in Eq. (\ref{eq:H_Ising}), a spin interacts with up to the third NN ones, the local energies relevant to the four spins are all the same. Now, we consider the case where the 4u single tetrahedron is replaced with a 3u1d one. Because all the four sites on the original 4u tetrahedron are energetically equivalent to one another, one can freely pick up one of the four sites for the down spin of the 3u1d tetrahedron. Thus, there are 4 different possible spin-configurations. In other words, the 3u1d tetrahedron interacting with outer tetrahedra possesses the same 4-fold degeneracy as that for the {\it isolated} one. Furthermore, as the extent of the interaction is within the third-NN spins, central small tetrahedra are not directly correlated to each other, so that all the 4u (4d) tetrahedra can be replaced with 3u1d (1u3d) ones independently. Noting that one cubic unit cell contains only one central small tetrahedron and each 3u1d or 1u3d central small tetrahedron has the 4-fold degeneracy, the total NOS in the type-II order is calculated as $4^{L^3} \times {\cal O}(1)$. The same counting method can also be applied to the type-III order in which the 4u and 4d tetrahedra in the original type-I order is replaced with the 2u2d ones. As the single 2u2d tetrahedron has 6-fold degeneracy, the total NOS in the type-III order is $6^{L^3} \times {\cal O}(1)$. With increasing the total number of spins $N=16L^3$, the NOS's in the type-II and type-III orders exponentially increase, suggestive of the massive degeneracy of the ground state in these orders. Given the total NOS, one obtain the ground-state entropy, namely, the residual entropy, per spin $s^R$ from the Boltzmann's entropy formula as $0$, $\frac{\ln(4)}{16}$, and $\frac{\ln(6)}{16}$ for the type-I, type-II, and type-III orders, respectively. Here, we have ignored the $\frac{\ln({\cal O}(1))}{16L^3}$ correction which vanishes in the thermodynamic limit of $L \rightarrow \infty$.

To verify the above argument, we calculate the entropy of the system $S(T)$ by means of MC simulations.
Since the classical Heisenberg spins exhibit a pathological behavior that the entropy diverges toward $T=0$, we employ instead to estimate the entropy the effective Ising model Eq. (\ref{eq:H_Ising}), bearing the collinear spin states in mind. Figure \ref{fig:entropy} shows the temperature dependences of the specific heat $C$, the averaged magnetic Bragg intensity $O^S_{(\frac{1}{2}\frac{1}{2}\frac{1}{2})}$, and the entropy per spin $S/(k_{\rm B}N)$ for the type-I (cyan), type-II (blue), and type-III (green) orders obtained for $L=4$. At low temperatures, we have performed $2 \times 10^5 - 2\times 10^7$ MC sweeps by using the temperature exchange method \cite{Fukushima_exchange}, the first half being discarded for thermalization. The entropy $S(T)$ is extracted from the relation $S(T)/(k_{\rm B}N)= \ln(2) - \int^{\infty}_T \frac{C(T')}{T'} \, dT' \, $ (for details, see Appendix B).

As one can see from the middle panel of Fig. \ref{fig:entropy}, in each case, $O^S_{(\frac{1}{2}\frac{1}{2}\frac{1}{2})}$ at the lowest temperature coincides with the corresponding value obtained in the original site-phonon Heisenberg model which is indicated by horizontal arrows in the middle panel of Fig. \ref{fig:entropy}, suggesting that the observed ordred phases are Ising analogs of the type-I, type-II, and type-III orders of the Heisenberg model. In each of the three cases, the entropy per spin $S(T)/(k_{\rm B}N)$ saturates to a constant value at temperatures lower than the sharp drop associated with the magnetic transition between the paramagnetic and $(\frac{1}{2},\frac{1}{2},\frac{1}{2})$ states. The saturation values, namely, the residual entropies, in the three cases are close to 0, $\frac{\ln(4)}{16}$, and $\frac{\ln(6)}{16}$, in good agreement with the analytical estimates in the type-I, type-II, and type-III orders, respectively. 
We note that in the type-III order, the specific heat $C(T)$ and the entropy $S(T)$ respectively exhibit a small bump and an associated weak kink near $T/J\simeq 0.6$ above the magnetic transition at $T/J=0.16$ (see Fig. \ref{fig:entropy}). This is associated with the development of the local spin correlation on the small tetrahedra, namely, the formation of the 2u2d configuration on the small tetrahedra having the relatively strong NN interaction $J_1^{\rm eff}=J(1-b)$ compared with that for the large tetrahedra $J_{1'}^{\rm eff}=J'(1-b')$.

We note in passing that concerning the type-I $(\frac{1}{2},\frac{1}{2},\frac{1}{2})$ order, the same magnetic order is reported in the spin-ice Kondo-lattice model where our `up' and `down' spins correspond to `in' and `out' spins \cite{Ice_Ishizuka_11} and in the $J_1$-$J_2$-$J_3$ Heisenberg model with $J_2=J_3$ \cite{J1J2J3_Mizoguchi_18}. A common feature of the three models including the present site-phonon model is the existence of a relatively strong third-neighbor antiferromagnetic interaction. An additional breathing bond-alternation may induce the type-II and type-III orders in the above two different models, but we will leave this issue for our future work.

\section{Summary and Discussions}
In this paper, the effects of lattice distortions on the spin ordering have been investigated in the classical antiferromagnetic Heisenberg model on the breathing pyrochlore lattice, where the breathing alternation is quantified by $J'/J$. Our MC simulations show that the SLC originating from the site-phonons induces a first-order transition into four different types of collinear spin states. One is characterized by the $(1,1,0)$ magnetic Bragg peaks at weaker SLC, and the other three by the $(\frac{1}{2},\frac{1}{2},\frac{1}{2})$ ones at stronger SLC.
In the weak SLC regime, the $(1,1,0)$ spin state is tetragonal-symmetric and is robust against the breathing lattice-distortions, i.e., the decrease in $J'/J$. In the strong SLC regime, the $(\frac{1}{2},\frac{1}{2},\frac{1}{2})$ spin state is cubic-symmetric down to moderate $J'/J$, but with further decreasing $J'/J$, it becomes non-cubic with its magnetic Bragg reflections almost unchanged. The non-cubic spin state peculiar to the breathing pyrochlores takes two different ordering patterns depending on the value of $J'/J$ and the strength of the SLC. We have demonstrated with use of the corresponding effective Ising model that the residual entropy associated with these two non-cubic orderings takes characteristic values of $\simeq \frac{\ln(4)}{16}k_{\rm B}$ and $\simeq \frac{\ln(6)}{16}k_{\rm B}$ per spin. 

In the type-II and type-III orders, the framework of the magnetic structure is long-range-ordered being reflected in the $(\frac{1}{2},\frac{1}{2},\frac{1}{2})$ magnetic Bragg peaks, whereas the detailed spin configuration in each small tetrahedron at the center of the cubic unit cell remains locally degenerate resulting in the residual entropy. In this respect, the type-II and type-III orders are partially ordered states. We emphasize again that such an emergent phenomenon is induced by the breathing bond-alternation of the pyrochlore lattice.

Now, we will discuss experimental implications of our results. For the uniform pyrochlore lattice, we have already argued that the effect of the site phonon is relevant to the spin-lattice-coupled ordering in chromium spinel oxides \cite{Site_AK_16}. For the breathing pyrochlore lattices, the $(1,1,0)$ state obtained at weaker SLC in the present model has the same spin structure as that in the low-temperature ordered phase of the tetragonal domain in the breathing pyrochlore antiferromagnets Li(In, Ga)Cr$_4$O$_8$ \cite{BrPyro_Nilsen_15, BrPyro_Saha_16}. The observed first-order nature of the magnetic transition in LiGaCr$_4$O$_8$ with the weak bond-alternation is in agreement with our result. This suggests that the present site-phonon Heisenberg model captures the essential part of the ordering mechanism not only of uniform but also of breathing pyrochlore antiferromagnets. Although in contrast to our result, the magnetic transition is of second order in LiInCr$_4$O$_8$, being separated by the structural one \cite{BrPyro_Tanaka_14,BrPyro_Nilsen_15,BrPyro_Saha_16,BrPyro_Lee_16}, this might be due to quantum effects specific to this compound with the strong bond-alternation. Indeed, the singlet formation has been reported at slightly higher temperatures \cite{BrPyro_Tanaka_14, BrPyro_Nilsen_15, BrPyro_Saha_16, BrPyro_Lee_16} like in the case of the quantum breathing pyrochlore antiferromagnet Ba$_3$Yb$_2$Zn$_5$O$_{11}$ having almost isolated tetrahedra \cite{qBrPyro_Kimura_14, qBrPyro_Haku_prb16, qBrPyro_Haku_jpsj16, qBrPyro_Rau_16}. Nevertheless, the SLC in LiInCr$_4$O$_8$ would be a key ingredient for the magnetic long-range ordering at lower temperatures, as is also suggested from the observation of the half-magnetization plateau \cite{BrPyro_Hdep_Okamoto_17, BrPyro_Hdep_Gen_18} which in the uniform case, is considered to be a manifestation of the SLC \cite{ZnCrO_Miyata_jpsj_11, ZnCrO_Miyata_prl_11, CdCrO_Kojima_08, HgCrO_Ueda_06,Bond_Penc_04, Bond_Motome_06, Bond_Shannon_10}.    

The $(\frac{1}{2},\frac{1}{2},\frac{1}{2})$ magnetic Bragg reflections, on the other hand, have not yet been observed experimentally not only in the breathing case but also in the uniform case \cite{StrongSLC_exp}. Although it might be difficult to significantly enhance the strength of both SLC's $b$ and $b'$, strong $b'$ may be possible if the breathing alternation is strong enough, because the relation $b'\propto 1/J'$ holds. In particular, the type-III non-cubic $(\frac{1}{2},\frac{1}{2},\frac{1}{2})$ order might be relatively easy to access because it can be stabilized in the relatively small $b$ and large $b'$ region if $J'/J$ is sufficiently small. 
If such material parameters are realized, collinearly aligned Heisenberg spins may freeze deep inside the type-III $(\frac{1}{2},\frac{1}{2},\frac{1}{2})$ ordered phase at lower temperatures. Such a spin freezing would result in the distinct macroscopic residual entropy estimated from the corresponding Ising model, $\simeq \frac{\ln(6)}{16}k_{\rm B}$ per spin. The observation of such residual entropy could be a definitive signature of the characteristic ordering, similarly to the Pauling entropy in the spin-ice compounds Dy$_2$Ti$_2$O$_7$ and Ho$_2$Ti$_2$O$_7$\cite{Ho2Ti2O7_Harris_97, Dy2Ti2O7_Ramirez_99,Ho2Ti2O7_Bramwell_01, Ho2Ti2O7_Gardner_01}.

\begin{acknowledgments}
The authors thank Y. Tanaka and M. Gen for useful discussions. We are thankful to ISSP, the University of Tokyo for providing us with CPU time. This work is supported by JSPS KAKENHI Grant Numbers JP16K17748, JP17H06137.
\end{acknowledgments}

\appendix
\section{$J({\bf q})$ for the effective Ising model Eq. (\ref{eq:H_Ising})}
Via the Fourier transform $\sigma_i=\sum_{\bf q}\sigma^a_{\bf q}\exp(i{\bf q}\cdot {\bf r}_i)$, we can rewrite the Ising Hamiltonian Eq. (\ref{eq:H_Ising}) into the following form (the notations are the same as those in the main text) \cite{Reimers_MF, Okubo_pyro}
\begin{eqnarray}
{\cal H}^{\rm eff}_{\rm Ising}-E_0 &=& \frac{1}{2}\sum_{i,j} J_{ij} \sigma_i \sigma_j =  \frac{N}{8}\sum_{\bf q} \sum_{a,b=1}^4 J^{ab}({\bf q}) \sigma^a_{\bf q} \sigma^b_{-{\bf q}} \nonumber\\
J^{11}({\bf q}) &=& 2 J_3^{\rm eff} \Big[ \cos\Big(\frac{q_x+q_y}{2}\Big) + \cos\Big(\frac{q_z-q_x}{2}\Big) \nonumber\\
&& \qquad + \cos\Big(\frac{q_y-q_z}{2}\Big) \Big] \nonumber\\
J^{22}({\bf q}) &=& 2 J_3^{\rm eff} \Big[ \cos\Big(\frac{q_x+q_y}{2}\Big) + \cos\Big(\frac{q_y+q_z}{2}\Big) \nonumber\\
&& \qquad + \cos\Big(\frac{q_x+q_z}{2}\Big) \Big] \nonumber\\
J^{33}({\bf q}) &=& 2 J_3^{\rm eff} \Big[ \cos\Big(\frac{q_z-q_x}{2}\Big) + \cos\Big(\frac{q_x-q_y}{2}\Big) \nonumber\\
&& \qquad + \cos\Big(\frac{q_y+q_z}{2}\Big) \Big] \nonumber\\
J^{44}({\bf q}) &=& 2 J_3^{\rm eff} \Big[ \cos\Big(\frac{q_y-q_z}{2}\Big) + \cos\Big(\frac{q_x-q_y}{2}\Big) \nonumber\\
&& \qquad + \cos\Big(\frac{q_x+q_z}{2}\Big) \Big] \nonumber\\
J^{12}({\bf q}) &=& \big[ J^{21}({\bf q}) \big]^\ast = J_{1}^{\rm eff} e^{i\frac{1}{4}(q_x+q_y)}+J_{1'}^{\rm eff} e^{-i\frac{1}{4}(q_x+q_y)} \nonumber\\
&& \qquad \qquad \, \, + \, 4 J_2^{\rm eff} \cos\Big(\frac{q_z}{2}\Big) \cos\Big(\frac{q_x-q_y}{4}\Big) \nonumber\\
J^{13}({\bf q}) &=& \big[ J^{31}({\bf q}) \big]^\ast = J_{1}^{\rm eff} e^{i\frac{1}{4}(q_x-q_z)}+J_{1'}^{\rm eff} e^{-i\frac{1}{4}(q_x-q_z)} \nonumber\\
&& \qquad \qquad \, \, + \, 4 J_2^{\rm eff} \cos\Big(\frac{q_y}{2}\Big) \cos\Big(\frac{q_x+q_z}{4}\Big)  \nonumber\\
J^{14}({\bf q}) &=& \big[ J^{41}({\bf q}) \big]^\ast = J_{1}^{\rm eff} e^{i\frac{1}{4}(q_y-q_z)}+J_{1'}^{\rm eff} e^{-i\frac{1}{4}(q_y-q_z)} \nonumber\\
&& \qquad \qquad \, \, + \, 4 J_2^{\rm eff} \cos\Big(\frac{q_x}{2}\Big) \cos\Big(\frac{q_y+q_z}{4}\Big)  \nonumber\\
J^{23}({\bf q}) &=& \big[ J^{32}({\bf q}) \big]^\ast = J_{1}^{\rm eff} e^{-i\frac{1}{4}(q_y+q_z)}+J_{1'}^{\rm eff} e^{i\frac{1}{4}(q_y+q_z)} \nonumber\\
&& \qquad \qquad \, \, + \, 4 J_2^{\rm eff} \cos\Big(\frac{q_x}{2}\Big) \cos\Big(\frac{q_y-q_z}{4}\Big)  \nonumber\\
J^{24}({\bf q}) &=& \big[ J^{42}({\bf q}) \big]^\ast = J_{1}^{\rm eff} e^{-i\frac{1}{4}(q_x+q_z)}+J_{1'}^{\rm eff} e^{i\frac{1}{4}(q_x+q_z)} \nonumber\\
&& \qquad \qquad \, \, + \, 4 J_2^{\rm eff} \cos\Big(\frac{q_y}{2}\Big) \cos\Big(\frac{q_x-q_z}{4}\Big)  \nonumber\\
J^{34}({\bf q}) &=& \big[ J^{43}({\bf q}) \big]^\ast = J_{1}^{\rm eff} e^{-i\frac{1}{4}(q_x-q_y)}+J_{1'}^{\rm eff} e^{i\frac{1}{4}(q_x-q_y)} \nonumber\\
&& \qquad \qquad \, \, + \, 4 J_2^{\rm eff} \cos\Big(\frac{q_z}{2}\Big) \cos\Big(\frac{q_x+q_y}{4}\Big)  . \nonumber\\
\end{eqnarray}
  
The lowest eigen value at ${\bf q}=\frac{2\pi}{a}(1,q,0)$, $J^{\rm min}(1,q,0)$, is calculated as
\begin{eqnarray}\label{eq:eigen_weak}
&&J^{\rm min}(1,q,0) \nonumber \\
&&= \left\{ \begin{array}{cc}
-(J_{1}^{\rm eff}+ J_{1'}^{\rm eff}) & \cdots ( J_{3}^{\rm eff}< J_{1}^{\rm eff}, \, J_{3}^{\rm eff} < J_{1'}^{\rm eff} ) \\
3J_{1'}^{\rm eff}- J_{1}^{\rm eff}-4J_{3}^{\rm eff} & \cdots ( J_{1'}^{\rm eff} < J_{3}^{\rm eff}, \, J_{1'}^{\rm eff} < J_{1}^{\rm eff}  ) \\
3J_{1}^{\rm eff}-J_{1'}^{\rm eff}-4J_{3}^{\rm eff}  & \cdots ( J_{1}^{\rm eff} < J_{3}^{\rm eff}, \,  J_{1}^{\rm eff} < J_{1'}^{\rm eff}  ) \\
\end{array} \right. \nonumber\\ ,
\end{eqnarray}
and the one at ${\bf q}=\frac{2\pi}{a}(\frac{1}{2},\frac{1}{2},\frac{1}{2})$, $J^{\rm min}(\frac{1}{2},\frac{1}{2},\frac{1}{2})$, is calculated as
\begin{eqnarray}\label{eq:eigen_strong}
 J^{\rm min}\Big(\frac{1}{2},\frac{1}{2},\frac{1}{2}\Big) &=& J_{1}^{\rm eff}+ J_{1'}^{\rm eff}-2J_{3}^{\rm eff} \nonumber\\
&-& \sqrt{3(J_{1}^{\rm eff}- J_{1'}^{\rm eff})^2 + (J_{1}^{\rm eff}+ J_{1'}^{\rm eff}+4J_{3}^{\rm eff})^2}. \nonumber\\
\end{eqnarray}

\section{Entropy calculation}

\begin{figure}[t]
\includegraphics[scale=0.50]{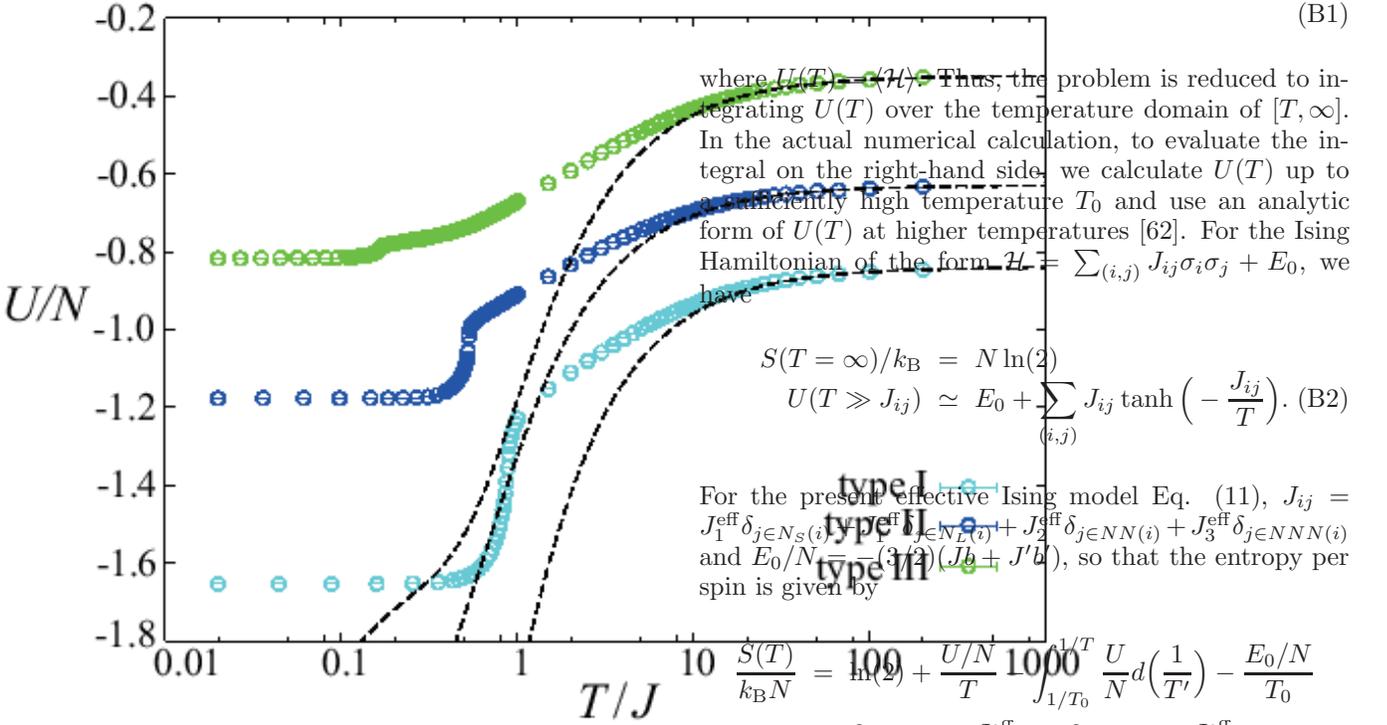}
\caption{The temperature dependence of the internal energy per spin, $U/N$, in the effective Ising model Eq. (\ref{eq:H_Ising}). $U$ is measured in units of $J$ and notations of the colored points are the same as those in Fig. \ref{fig:entropy}. Black dashed curves represent the analytic form of $U/N$ given by Eq. (\ref{eq:e_analytic}). \label{fig:e-app}}
\end{figure}

The entropy of the system is given by 
\begin{eqnarray}
S(T)/k_{\rm B} &=& \int^{T}_0 \Big[ \frac{dU(T')}{ d T'}\Big]\frac{dT'}{T'} \nonumber\\
&=& \frac{U(T)}{T} -\int^{1/T}_0 U(T') d\Big( \frac{1}{T'} \Big) + S(T=\infty) , \nonumber\\
\end{eqnarray}
where $U(T)=\langle {\cal H} \rangle$. Thus, the problem is reduced to integrating $U(T)$ over the temperature domain of $[T,\infty]$.
In the actual numerical calculation, to evaluate the integral on the right-hand side, we calculate $U(T)$ up to a sufficiently high temperature $T_0$ and use an analytic form of $U(T)$ at higher temperatures \cite{SG_Kirkpatrick}. For the Ising Hamiltonian of the form ${\cal H} = \sum_{(i,j)} J_{ij} \sigma_i \sigma_j + E_0$, we have
\begin{eqnarray}
S(T=\infty)/k_{\rm B} &=& N \ln (2) \nonumber\\
U(T \gg J_{ij}) &\simeq & E_0+\sum_{(i,j)} J_{ij}\tanh\Big( -\frac{J_{ij}}{T} \Big).
\end{eqnarray}
For the present effective Ising model Eq. (\ref{eq:H_Ising}), $J_{ij}=J_{1}^{\rm eff}\delta_{j\in N_S(i)}+J_{1'}^{\rm eff}\delta_{j\in N_L(i)}+J_{2}^{\rm eff}\delta_{j\in NN(i)}+J_{3}^{\rm eff}\delta_{j\in NNN(i)}$ and $E_0/N=-(3/2)(Jb+J'b')$, so that the entropy per spin is given by
\begin{eqnarray}
\frac{S(T)}{k_{\rm B}N} &=& \ln (2) + \frac{ U/N }{T} - \int^{1/T}_{1/T_0} \frac{U}{N} d\Big( \frac{1}{T'}\Big) - \frac{E_0/N}{T_0} \nonumber\\
&+& \frac{3}{2} \ln \cosh\Big( \frac{J_{1}^{\rm eff}}{T_0}\Big) + \frac{3}{2}\ln \cosh\Big( \frac{J_{1'}^{\rm eff}}{T_0}\Big) \nonumber\\
&& +6\ln \cosh\Big( \frac{J_{2}^{\rm eff}}{T_0}\Big)+3\ln \cosh\Big( \frac{J_{3}^{\rm eff}}{T_0}\Big) .
\end{eqnarray}
To determine the upper cutoff $T_0$, we compare the numerically obtained $U(T)$ curve to the analytically obtained one.  

As one can see from Fig. \ref{fig:e-app}, the high-temperature analytic form of the internal energy
\begin{eqnarray}\label{eq:e_analytic}
 \frac{U}{N} &\simeq& \frac{E_0}{N}-\frac{3}{2} J_{1}^{\rm eff} \tanh\Big( \frac{J_{1}^{\rm eff}}{T}\Big) - \frac{3}{2} J_{1'}^{\rm eff}\tanh\Big( \frac{J_{1'}^{\rm eff}}{T}\Big) \nonumber\\
&-& 6J_2^{\rm eff} \tanh\Big( \frac{J_2^{\rm eff}}{T} \Big)-3J_3^{\rm eff} \tanh\Big( \frac{J_3^{\rm eff}}{T}\Big)
\end{eqnarray}
is in good agreement with the numerically obtained temperature dependence of $U/N$ down to $T/J \simeq 50$. Thus, we take the upper cutoff for the numerical calculation to be $T_0/J=200$.


\begin{thebibliography}{50}
\bibitem{ZnCrO_Lee_00} S.-H. Lee, C. Broholm, T. H. Kim, W. Ratcliff II, and S-W. Cheong, Phys. Rev. Lett. {\bf 84}, 3718 (2000).
\bibitem{CdCrO_Chung_05} J.-H. Chung, M. Matsuda, S.-H. Lee, K. Kakurai, H. Ueda, T. J. Sato, H. Takagi, K.-P. Hong, and S. Park, Phys. Rev. Lett. {\bf 95}, 247204 (2005).
\bibitem{HgCrO_Ueda_06} H. Ueda, H. Mitamura, T. Goto, and Y. Ueda, Phys. Rev. B {\bf 73}, 094415 (2006).
\bibitem{MgCrO_Ortega_08} L. Ortega-San-Martin, A. J. Williams, C. D. Gordon, S. Klemme, and J. P. Attfield, J. Phys. Condens. Matter {\bf 20}, 104238, (2008).
\bibitem{ZnCrO_Lee_08} S.-H. Lee, W. Ratcliff, Q. Huang, T. H. Kim, and S-W. Cheong, Phys. Rev. B {\bf 77}, 014405 (2008).
\bibitem{HgCrO_Matsuda_07} M. Matsuda, H. Ueda, A. Kikkawa, Y. Tanaka, K. Katsumata, Y. Narumi, T. Inami, Y. Ueda, and S.-H. Lee, Nat. Phys. {\bf 3}, 397 (2007).

\bibitem{BrPyro_Okamoto_13} Y. Okamoto, G. J. Nilsen, J. P. Attfield, and Z. Hiroi, Phys. Rev. Lett. {\bf 110}, 097203 (2013).
\bibitem{BrPyro_Tanaka_14} Y. Tanaka, M. Yoshida, M. Takigawa, Y. Okamoto, and Z. Hiroi, Phys. Rev. Lett. {\bf 113}, 227204 (2014).
\bibitem{BrPyro_Nilsen_15} G. J. Nilsen, Y. Okamoto, T. Masuda, J. Rodriguez-Carvajal, H. Mutka, T. Hansen, and Z. Hiroi, Phys. Rev. B {\bf 91}, 174435 (2015).
\bibitem{BrPyro_Saha_16} R. Saha, F. Fauth, M. Avdeev, P. Kayser, B. J. Kennedy, and A. Sundaresan, Phys. Rev. B {\bf 94}, 064420 (2016).
\bibitem{BrPyro_Lee_16} S. Lee, S.-H. Do, W.-J. Lee, Y. S. Choi, M. Lee, E. S. Choi, A. P. Reyes, P. L. Kuhns, A. Ozarowski, and K.-Y. Choi, Phys. Rev. B {\bf 93}, 174402 (2016).
\bibitem{BrPyro_Saha_17} R. Saha, R. Dhanya, C. Bellin, K. Beneut, A. Bhattacharyya, A. Shukla, C. Narayana, E. Suard, J. Rodriguez-Carvajal, and A. Sundaresan, Phys. Rev. B {\bf 96}, 214439 (2017).
\bibitem{Reimers_MC_92} J. N. Reimers, Phys. Rev. B {\bf 45}, 7287 (1992).
\bibitem{Moessner-Chalker_prl} R. Moessner and J. T. Chalker, Phys. Rev. Lett. {\bf 80}, 2929 (1998).
\bibitem{Moessner-Chalker_prb} R. Moessner and J. T. Chalker, Phys. Rev. B {\bf 58}, 12049 (1998).

\bibitem{BrPyro_doped_Okamoto_15} Y. Okamoto, G. J. Nilsen, T. Nakazono, and Z. Hiroi, J. Phys. Soc. Jpn. {\bf 84}, 043707 (2015).
\bibitem{BrPyro_doped_Wang_17} D. Wang, C. Tan, K. Huang, and L. Shu, Chinese Phys. Lett. {\bf 33}, 127501 (2016).  
\bibitem{BrPyro_doped_Wawrzynczak_17} R. Wawrzynczak, Y. Tanaka, M. Yoshida, Y. Okamoto, P. Manuel, N. Casati, Z. Hiroi, M. Takigawa, and G. J. Nilsen, Phys. Rev. Lett. {\bf 119}, 087201 (2017).
\bibitem{BrPyro_Sulfides_Okamoto_18} Y. Okamoto, M. Mori, N. Katayama, A. Miyake, M. Tokunaga, A. Matsuo, K. Kindo, and K. Takenaka, J. Phys. Soc. Jpn. {\bf 87}, 034709 (2018).
\bibitem{BrPyro_Sulfides_Pokharel_18} G. Pokharel, A. F. May, D. S. Parker, S. Calder, G. Ehlers, A. Huq, S. A. J. Kimber, H. S. Arachchige, L. Poudel, M. A. McGuire, D. Mandrus, and A. D. Christianson, Phys. Rev. B {\bf 97}, 134117 (2018).
\bibitem{qBrPyro_Kimura_14} K. Kimura, S. Nakatsuji, and T. Kimura, Phys. Rev. B {\bf 90}, 060414(R) (2014).  
\bibitem{qBrPyro_Haku_prb16} T. Haku, K. Kimura, Y. Matsumoto, M. Soda, M. Sera, D. Yu, R. A. Mole, T. Takeuchi, S. Nakatsuji, Y. Kono, T. Sakakibara, L.-J. Chang, and T. Masuda, Phys. Rev. B {\bf 93}, 220407(R) (2016).  
\bibitem{qBrPyro_Haku_jpsj16} T. Haku, M. Soda, M. Sera, K. Kimura, S. Itoh, T. Yokoo, and T. Tasuda, J. Phys. Soc. Jpn. {\bf 85}, 034721 (2016).  
\bibitem{qBrPyro_Rau_16} J. G. Rau, L. S. Wu, A. F. May, L. Poudel, B. Winn, V. O. Garlea, A. Huq, P. Whitfield, A. E. Taylor, M. D. Lumsden, M. J. P. Gingras, and A. D. Christianson, Phys. Rev. Lett. {\bf 116}, 257204 (2016).

\bibitem{BrPyro_NNmodel_Benton_15} O. Benton and N. Shannon, J. Phys. Soc. Jpn. {\bf 84}, 104710 (2015).
\bibitem{Pyro_NNmodel_Tsunetsugu_00} H. Tsunetsugu, J. Phys. Soc. Jpn. {\bf 70}, 640 (2000).
\bibitem{BrPyro_NNmodel_Tsunetsugu_17} H. Tsunetsugu, Prog. Theor. Exp. Phys. {\bf 3}, 033I01 (2017). 

\bibitem{SLC_Tchernyshyov_prl_02} O. Tchernyshyov, R. Moessner, and S. L. Sondhi, Phys. Rev. Lett. {\bf 88}, 067203 (2002).
\bibitem{SLC_Tchernyshyov_prb_02} O. Tchernyshyov, R. Moessner, and S. L. Sondhi, Phys. Rev. B {\bf 66}, 064403 (2002).
\bibitem{SLC_Yamashita_00} Y. Yamashita and K. Ueda, Phys. Rev. Lett. {\bf 85}, 4960 (2000).
\bibitem{Bond_Penc_04} K. Penc, N. Shannon, and H. Shiba, Phys. Rev. Lett. {\bf 93}, 197203 (2004).
\bibitem{Bond_Motome_06} Y. Motome, K. Penc, and N. Shannon, J. Magn. Magn. Mater. {\bf 300}, 57 (2006).
\bibitem{Bond_Shannon_10} N. Shannon, K. Penc, and Y. Motome, Phys. Rev. B {\bf 81}, 184409 (2010).
\bibitem{Site_Jia_05} C. Jia, J. H. Nam, J. S. Kim, and J. H. Han, Phys. Rev. B {\bf 71}, 212406 (2005). 
\bibitem{Site_Bergman_06} D. L. Bergman, R. Shindou, G. A. Fiete, and L. Balents, Phys. Rev. B {\bf 74}, 134409 (2006).
\bibitem{Site_Wang_08} F. Wang and A. Vishwanath, Phys. Rev. Lett. {\bf 100}, 077201 (2008). 
\bibitem{Site_AK_16} K. Aoyama and H. Kawamura, Phys. Rev. Lett. {\bf 116}, 257201 (2016).  

\bibitem{CdCrO_Kojima_08} E. Kojima, A. Miyata, S. Miyabe, S. Takeyama, H. Ueda, and Y. Ueda, Phys. Rev. B {\bf 77}, 212408 (2008).
\bibitem{ZnCrO_Miyata_jpsj_11} A. Miyata, H. Ueda, Y. Ueda, Y. Motome, N. Shannon, K. Penc, and S. Takeyama, J. Phys. Soc. Jpn. {\bf 80}, 074709 (2011).
\bibitem{ZnCrO_Miyata_prl_11} A. Miyata, H. Ueda, Y. Ueda, H. Sawabe, and S. Takeyama, Phys. Rev. Lett. {\bf 107}, 207203 (2011).
\bibitem{ZnCrO_Miyata_jpsj_12} A. Miyata, H. Ueda, Y. Ueda, Y. Motome, N. Shannon, K. Penc, and S. Takeyama, J. Phys. Soc. Jpn. {\bf 81}, 114701 (2012).
\bibitem{SLC_Chern_06} Gia-Wei Chern, C. J. Fennie, and O. Tchernyshyov, Phys. Rev. B {\bf 74}, 060405(R) (2006).

\bibitem{ZnCrO_Sushkov_05} A. B. Sushkov, O. Tchernyshyov, W. Ratcliff, S. W. Cheong, and H. D. Drew, Phys. Rev. Lett. {\bf 94}, 137202 (2005).
\bibitem{ZnCrO_Fennie_06} C. J. Fennie and K. M. Rabe, Phys. Rev. Lett. {\bf 96}, 205505 (2006).
\bibitem{CdCrO_Rudolf_07} T. Rudolf, Ch. Kant, F. Mayr, J. Hemberger, V. Tsurkan and A. Loidl, New J. Phys. {\bf 9}, 76 (2007).
\bibitem{CdCrO_Aguilar_08} R. Valdes Aguilar, A. B. Sushkov, Y. J. Choi, S.-W. Cheong, and H. D. Drew, Phys. Rev. B {\bf 77}, 092412 (2008).
\bibitem{CdZn-CrO_Kant_09} Ch. Kant, J. Deisenhofer, T. Rudolf, F. Mayr, F. Schrettle, A. Loidl, V. Gnezdilov, D. Wulferding, P. Lemmens, and V. Tsurkan, Phys. Rev. B {\bf 80}, 214417 (2009).
\bibitem{CdCrO_Bhattacharjee_11} S. Bhattacharjee, S. Zherlitsyn, O. Chiatti, A. Sytcheva, J. Wosnitza, R. Moessner, M. E. Zhitomirsky, P. Lemmens, V. Tsurkan, and A. Loidl, Phys. Rev. B {\bf 83}, 184421 (2011).

\bibitem{Reimers_MF} J. N. Reimers, A. J. Berlinsky, and A. C. Shi, Phys. Rev. B {\bf 43}, 865 (1991). 
\bibitem{Okubo_pyro} T. Okubo, T. H. Nguyen, and H. Kawamura, Phys. Rev. B {\bf 84}, 144432 (2011).
\bibitem{Loop_Shinaoka_14} H. Shinaoka, Y. Tomita, and Y. Motome, Phys. Rev. B {\bf 90}, 165119 (2014).
\bibitem{Fukushima_exchange} K. Fukushima and K. Nemoto, J. Phys. Soc. Jpn. {\bf 65}, 1604 (1996).

\bibitem{Ice_Ishizuka_11} H. Ishizuka, M. Udagawa, and Y. Motome, J. Phys. Soc. Jpn. {\bf 81}, 113706 (2011).
\bibitem{J1J2J3_Mizoguchi_18} T. Mizoguchi, L. D. C. Jaubert, R. Moessner, and M. Udagawa, Phys. Rev. B {\bf 98}, 144446 (2018).


\bibitem{BrPyro_Hdep_Okamoto_17} Y. Okamoto, D. Nakamura, A. Miyake, S. Takeyama, M. Tokunaga, A. Matsuo, K. Kindo, and Z. Hiroi, Phys. Rev. B {\bf 95}, 134438 (2017).
\bibitem{BrPyro_Hdep_Gen_18} M. Gen, D. Nakamura, Y. Okamoto, and S. Takeyama, J. Magn. Magn. Mater. {\bf 473}, 387 (2019).

\bibitem{StrongSLC_exp} Although Bragg reflections at $(\frac{1}{2},\frac{1}{2},\frac{1}{2})$ have been reported as a magnetic domain in the uniform pyrochlore antiferromagnet ZnCr$_2$O$_4$ \cite{ZnCrO_Lee_08}, the experimentally proposed spin structure is different from the one obtained in the strong SLC regime of the site-phonon Heisenberg model. 

\bibitem{Ho2Ti2O7_Harris_97} M. J. Harris, S. T. Bramwell, D. F. McMorrow, T. Zeiske, and K. W. Godfrey, Phys. Rev. Lett. {\bf 79}, 2554 (1997).
\bibitem{Dy2Ti2O7_Ramirez_99} A. P. Ramirez, A. Hayashi, R. J. Cava, R. Siddharthan, and B. S. Shastry, nature {\bf 399}, 333 (1999).
\bibitem{Ho2Ti2O7_Bramwell_01} S. T. Bramwell, M. J. Harris, B. C. den Hertog, M. J. P. Gingras, J. S. Gardner, D. F. McMorrow, A. R. Wildes, A. L. Cornelius, J. D. M. Champion, R. G. Melko, and T. Fennell, Phys. Rev. Lett. {\bf 87}, 047205 (2001).
\bibitem{Ho2Ti2O7_Gardner_01} A. L. Cornelius and J. S. Gardner, Phys. Rev. B {\bf 64}, 060406(R) (2001).

\bibitem{SG_Kirkpatrick} S. Kirkpatrick, Phys. Rev. B {\bf 16}, 4630 (1977).
\end{thebibliography}
\end{document}